\documentclass[aps,pre,twocolumn,amsmath,amssymb,superscriptaddress]{revtex4-2}
\usepackage[colorlinks]{hyperref}
\usepackage{graphicx}
\usepackage[dvipsnames]{xcolor}
\usepackage{braket}
\usepackage{soul}
\usepackage{bbm}
\usepackage{url}

\newcommand{\im}{{\rm i}}
\newcommand{\Ham}{\mathcal{H}}
\newcommand{\SE}{Schr\"odinger }
\DeclareMathOperator\Arg{Arg}

\definecolor{blue1}{rgb}{0 0.4470 0.7410}
\definecolor{green1}{rgb}{0.4660 0.6740 0.1880}
\definecolor{red1}{rgb}{0.8500 0.3250 0.0980}
\definecolor{purple1}{rgb}{0.4940 0.1840 0.5560}
\definecolor{black1}{rgb}{0 0 0}
\definecolor{green2}{rgb}{0.7020    0.9137    0.6784}
\definecolor{green3}{rgb}{0.2275    0.7333    0.2667}
\definecolor{green4}{rgb}{0.1600    0.3800    0.2700}
\definecolor{blue2}{rgb}{0.6627    0.9529    0.9922}
\definecolor{blue3}{rgb}{0.1647    0.6000    0.8431}
\definecolor{blue4}{rgb}{0.1294    0.1961    0.8431}
\definecolor{yellow1}{rgb}{0.9290  0.6940    0.1250}

\newcommand{\sutdphys}{Science, Mathematics and Technology Cluster, Singapore
University of Technology and Design, 8 Somapah Road, 487372 Singapore}
\newcommand{\sutdepd}{EPD Pillar, Singapore University of Technology and Design, 8 Somapah Road, 487372 Singapore}

\newcommand{\cqt}{Centre for Quantum Technologies, National University of Singapore 117543, Singapore} 
\newcommand{\majulab}{MajuLab, CNRS-UNS-NUS-NTU International Joint Research Unit, UMI 3654, Singapore}  
\newcommand{\sichuan}{College of Physics and Electronic Engineering, and Center for Computational Sciences, Sichuan Normal University, Chengdu 610068, China}

\begin{document}
	
\title{Paths towards time evolution with larger neural-network quantum states}

\author{Wenxuan Zhang} 
\affiliation{\cqt} 
\affiliation{\sutdphys}
\author{Bo Xing}   
\affiliation{\sutdphys}
\author{Xiansong Xu} 
\affiliation{\sutdphys}
\affiliation{\sichuan}

\author{Dario Poletti} 
\affiliation{\sutdphys}
\affiliation{\sutdepd}
\affiliation{\cqt}
\affiliation{\majulab}

\begin{abstract}  
In recent years, the neural-network quantum states method has been investigated to study the ground state and the time evolution of many-body quantum systems. 
Here we expand on the investigation and consider a quantum quench from the paramagnetic to the anti-ferromagnetic phase in the tilted Ising model. We use two types of neural networks, a restricted Boltzmann machine and a feed-forward neural network. We show that for both types of networks, the projected time-dependent variational Monte Carlo (p-tVMC) method performs better than the non-projected approach. 
We further demonstrate that one can use K-FAC or minSR in conjunction with p-tVMC to reduce the computational complexity of the stochastic reconfiguration approach, thus allowing the use of these techniques for neural networks with more parameters. 

\end{abstract}

\date{\today}

\maketitle

\section{Introduction} 
The rapid development of machine learning and neural networks has offered fresh insights and opportunities in natural sciences. Being a highly nonlinear method that is designed to deal with big data, neural networks could potentially help to tackle exponential complexity in quantum many-body physics. The seminal work by Carleo and Troyer proposed the use of neural networks as an ansatz~\cite{carleo2017solving}, known as the neural-network quantum states (NNQS), to study the ground state and dynamics of many-body quantum systems.  
Since then, NNQS have been applied on a variety of fields, such as the study of the ground state of many-body quantum systems, including molecules and frustrated Hamiltonians~\cite{choo2019two, choo2020fermionic,  hermann2020deep, guo2021scheme, ren2023towards, han2019solving, pfau2020ab, westerhout2020generalization, roth2023high, szabo2020neural, chen2022neural}, unitary time evolution~\cite{schmitt2020quantum, gutierrez2022real,  SinibaldiVicentini2023, donatella2023dynamics, burau2021unitary, SchmittZurek2022} and even dissipative systems~\cite{vicentini2019variational, nagy2019variational, yoshioka2019constructing, hartmann2019neural, reh2021time}. 
Explorations and progress in this direction are also documented in numerous reviews~\cite{carleo2019machine, jia2019quantum, carrasquilla2020machine, guest2018deep, larkoski2020jet}.    
In these explorations, different neural network models have been considered, including restricted Boltzmann machines (RBM)~\cite{smolensky1986information, nomura2021purifying, melko2019restricted, golubeva2022pruning, zen2020transfer,  park2022expressive, lu2019efficient, park2020geometry, nomura2017restricted, nomura2021helping, deng2017quantum, kingamin2024}, feed-forward neural networks (FNN)~\cite{choo2018symmetries, cai2018approximating, saito2017solving, beer2020training, inui2021determinant}, convolutional neural networks~\cite{gutierrez2022real, yang2020deep, irikura2020neural, vieijra2021many, liang2018solving, liu2021random, saito2018machine, roth2021group, fu2024lattice, reh2023optimizing, herrera2021convolutional}, recurrent neural networks \cite{kingamin2024}, and autoregressive networks~\cite{sharir2020deep, luo2022autoregressive} such as transformers~\cite{zhang2023transformer}. 

Key aspects of using neural networks to represent many-body quantum systems are their representative power and their trainability. For NNQS, stochastic reconfiguration (SR)\cite{Sorella1998} has emerged as a method of choice to train the networks. However, the use of NNQS for larger many-body quantum systems requires neural networks with more parameters, thus making the use of SR more challenging. For ground-state search, different tools have been developed, from doing SR optimization on the fly \cite{vicentini2022netket}, Kronecker-factored approximate curvature (K-FAC) \cite{martens2015optimizing}, sequential local optimization (SLO) \cite{zhang2023ground} and minSR \cite{chen2023efficient, rende2023simple}. However, these approaches have not been extended and tested for time evolution investigations.      

An important dynamical scenario to investigate further is the Hamiltonian quench: a system prepared in the ground state of a Hamiltonian experiences a sudden change in its Hamiltonian. Such quenches are frequently used to study the stability of the prepared state and play important roles in the study of thermalization~\cite{kollath2007quench, polkovnikov2011colloquium, yukalov2011equilibration, cassidy2011generalized, canovi2011quantum, rigol2009quantum}, many-body localization~\cite{serbyn2014quantum, schreiber2015observation, kjall2014many, bardarson2012unbounded, vosk2014dynamical}, and many-body scars~\cite{serbyn2021quantum, su2023observation, schecter2019weak, ho2019periodic, lin2020quantum}. In particular, it was shown that one can observe dynamical phase transitions as the dynamics can present singular behavior for certain protocols and Hamiltonians~\cite{heyl2013dynamical}. Hamiltonian quenches can be particularly hard to study because, while the initial state can be readily prepared to have zero or low entanglement entropy, the bipartite entanglement entropy of the system generally grows linearly in time after the quench. These systems are described by volume-law states, i.e. states in which the bipartite entanglement entropy grows linearly in time and saturates at a value proportional to the system size~\cite{cincio2007entropy, pastori2019disentangling}. 

In this work, we consider a quench from the paramagnetic to the antiferromagnetic phase in the non-integrable 1D tilted Ising model.
We analyze the roles of the evolution algorithm, optimization routine, neural network size, and number of samples in influencing the predicted quench dynamics. Most importantly, we infer the generic trends by investigating different neural networks: FNN and RBM. 
We experiment with two different types of evolution algorithms, namely the time-dependent variational Monte Carlo algorithm (tVMC)~\cite{carleo2017solving,schmitt2020quantum, rende2023simple, lee2021neural} and the projected time-dependent variational Monte Carlo algorithm (p-tVMC)~\cite{sinibaldi2023unbiasing, gutierrez2022real, donatella2023dynamics, burau2021unitary}. The former is a direct time-evolution of the neural network parameters derived from the Schr\"odinger equation, while the latter revolves around the optimization of a loss function built from the concepts of wave function overlap and infidelity. In our investigation, we find that p-tVMC performs better than tVMC in predicting the dynamics ensuing the quench. 
Furthermore, p-tVMC is readily amenable to the optimization approaches developed for ground state search. 
Here we show that, for larger system sizes, different methods may be adopted to deal with the large number of network parameters. One solution is to divide the network into portions and optimize each portion sequentially, similar to K-FAC. We also show that one can use minSR, which allows us to increase the number of network parameters without significantly impeding the computational efficiency of the numerical simulation.          

The manuscript is organized in the following manner: in Sec.~\ref{sec_NNQS}, we introduce the NNQS we use; in Sec.~\ref{sec_HamQuench}, we describe the Hamiltonian we study; in Sec.~\ref{sec_OptApproaches}, we present the two different approaches used to approximate the time evolution and give details about minSR and K-FAC; in Sec.~\ref{sec_figuresMerit}, we describe the quantities used to characterize the performance of our algorithms; we present the key results in Sec.~\ref{sec_results} and draw our conclusions in Sec.~\ref{sec_conclusions}.

\section{Neural Network Quantum States Models Considered and Their Characterization }\label{sec_NNQS}
We wish to analyze the roles of different hyperparameters in training neural networks, e.g. the number of parameters and samples, the optimization routine, etc. To uncover the generic trends, we explore two types of neural network architectures, FNN and RBM. 

In the FNN, the input layer receives the physical configurations $\pmb{x}=\left(x_1, x_2, ..., x_L\right)$, where $x_l \in \{-1, 1\}$. 
Every hidden node in the $k$-th layer, denoted by $\pmb{u}^{[k]}$, is connected to all the nodes of the previous $(k-1)$-th layer via the weight matrix $\pmb{W}^{[k]}$ and the local bias $\pmb{b}^{[k]}$. The values of the $k$-th layer are thus written as 
\begin{equation}\label{fnn_eq}
	\pmb{u}^{[k]} = \pmb{W}^{[k]} f(\pmb{u}^{[k-1]}) + \pmb{b}^{[k]}
\end{equation} 
where $\{\pmb{b}^{[k]}, \pmb{W}^{[k]}\} \in \pmb{\theta}$ are the variational parameters in the $k$-th layer. We use $\pmb{\theta}$ to represent all the network's parameters, and $\pmb{f}$ is a non-linear activation function. 
For the activation function, we use the Taylor expansion of the logarithm of hyperbolic cosine function $\ln\cosh(\pmb{x})$~\cite{schmitt2020quantum},  
\begin{equation}\label{act_func}
	f(\pmb{u}) = \pmb{u} - \frac{1}{3}\pmb{u}^3 + \frac{2}{15}\pmb{u}^5
.\end{equation} 
The unnormalized output is associated with the logarithm of the probability amplitude. For a particular configuration $\pmb{x}$, 
\begin{equation}\label{psi_eq_FNN}
	\psi(\pmb{x};\pmb{\theta})_{\mathrm{FNN}} = \exp\left(\pmb{u}^{[K]}\right)
\end{equation}
where the index $K$ refers to the last layer of the FNN. 

On the other hand, the RBM ansatz with $H$ hidden nodes returns the probability amplitude of the wave function as 
\begin{equation}\label{psi_eq_RBM} 
	\psi(\pmb{x};\pmb{\theta})_{\mathrm{RBM}} = \exp\left(\sum_l a_lx_l\right) \prod_{h=1}^H 2\cosh\left(b_h + \sum_{l=1}^L W_{h,l}x_l\right)
\end{equation}
where $\{\pmb{a}, \pmb{b},\pmb{W}\}\in \pmb{\theta}$ are the variational parameters of the RBM.  

We rely on the Monte Carlo method~\cite{metropolis1949,metropolis1953,hastings1970} to obtain the expectation values $\langle O \rangle$ from the neural network representation of the wave function. For operators that are diagonal in the computational basis, we get 
\begin{align}
    \langle O \rangle &= \frac{\sum_{\pmb{x}} |\psi(\pmb{x};\pmb{\theta})|^2 O(\pmb{x})}{\sum_{\pmb{z}} |\psi(\pmb{z};\pmb{\theta})|^2} \approx \frac{1}{N_s} \sum_{n} O(\pmb{x}_n),  
\end{align}
where the configuration $\pmb{x}_n$ is sampled from the probability distribution $P(\pmb{x};\pmb{\theta}) = |\psi(\pmb{x};\pmb{\theta})|^2 / \left(\sum_{\pmb{z}} |\psi(\pmb{z};\pmb{\theta})|^2\right)$ and $N_s$ is the total number of samples.  

For operators $O$ that are not diagonal in the configuration basis, such as the energy operator, we use
\begin{align}
    \langle O \rangle & = \frac{\sum_{\pmb{x}, \pmb{y}} \psi(\pmb{x};\pmb{\theta})^*\;  O_{\pmb{x}, \pmb{y}} \;  \psi(\pmb{y};\pmb{\theta}) }{\sum_{\pmb{z}} |\psi(\pmb{z};\pmb{\theta})|^2} \nonumber \\  
    & = \sum_{\pmb{x}} \frac{|\psi(\pmb{x};\pmb{\theta})|^2}{\sum_{\pmb{z}} |\psi(\pmb{z};\pmb{\theta})|^2} O_{loc}(\pmb{x}) \\
    & \approx \frac{1}{N_s} \sum_{n} O_{loc}(\pmb{x}_n), \label{eq_Oobs}
\end{align}
with the ``local'' evaluation of the observable 
\begin{align}
    O_{loc}(\pmb{x}) = \sum_{ \pmb{y} } O_{\pmb{x}, \pmb{y}} \frac{\psi(\pmb{y};\pmb{\theta})}{\psi(\pmb{x};\pmb{\theta})}.  \label{eq_Oloc} 
\end{align}
Similarly, we draw the configurations $\pmb{x}$ from $P(\pmb{x};\pmb{\theta})$ via the Monte Carlo method. The summation over $\pmb{y}$ in Eq.~(\ref{eq_Oobs}) can be computed efficiently when $O$ is sparse. 

Additionally, we define a $N_p \times N_p$ geometric tensor $\pmb{S}$,
\begin{align}
    S_{m,n} =& \left\langle \left( \frac{\partial \ln[\psi(\pmb{x};\pmb{\theta})]}{\partial \theta_m} \right)^* \; \frac{\partial \ln[\psi(\pmb{x};\pmb{\theta})]}{\partial \theta_n} \right\rangle \nonumber \\ 
    &- \left\langle \frac{\partial \ln[\psi(\pmb{x};\pmb{\theta})]}{\partial \theta_m} \right\rangle^*  \left\langle \frac{\partial \ln[\psi(\pmb{x};\pmb{\theta})]}{\partial \theta_n} \right\rangle,
    \label{eq_GT}
\end{align}
where $N_p$ is the number of neural network parameters, 
Computationally, it is useful to define an $N_p \times N_s$  auxiliary matrix $\pmb{X}$ with elements 
\begin{align}\label{eq_x}
    X_{m,k} =& \frac{1}{\sqrt{N_s}}\left(\left(\frac{\partial \ln[\psi(\pmb{x}_k;\pmb{\theta})]}{\partial \theta_m} \right)^* - \left\langle \frac{\partial \ln[\psi(\pmb{x};\pmb{\theta})]}{\partial \theta_m} \right\rangle^* \right)  
\end{align}
where $\pmb{x}_k$ is the $k$-th samples. 
We can then write 
\begin{align}
    \pmb{S} = \pmb{X}\pmb{X}^\dagger. \label{eq:SandXX}
\end{align}
We highlight that the geometric tensor $\pmb{S}$ may not be well conditioned. We regularize it by introducing a small shift in the main diagonal $\pmb{S}_{\rm{reg}} = \pmb{S} + \lambda \mathbbm{1}$, where $\mathbbm{1}$ is the identity matrix and $\lambda$ is the regularization parameter.
For simplicity of expression, we drop the subscript and use just $\pmb{S}$ to represent the regularized geometric tensor. Lastly, we highlight that regardless of the type of neural network, we use the notation $\psi_t(\pmb{x})$ to indicate the probability amplitude of configuration $\pmb{x}$ at time $t$, with the network parameters $\pmb{\theta}(t)$. 


\section{Hamiltonian Analyzed}\label{sec_HamQuench} 
We focus on the 1D tilted Ising model with open boundary conditions
\begin{equation}\label{TIM}
	\Ham = J\sum_{l=1}^{L-1} \sigma_l^z \sigma_{l+1}^z - \sum_{l=1}^L \left( h_x\sigma_l^x + h_z \sigma_l^{z} \right),
\end{equation}
where $h_x$ and $h_z$ are the strengths of the local fields along the $x$ and $z$ directions, $J$ is the coupling constant between nearest-neighbors spins, and $\sigma_l^a$ with $a = x,y,z$ are the Pauli matrices at the site $l$. When $J, \;h_x$, and $h_z$ are non-zero, the model is interacting and non-integrable. For $J>0$, stronger local fields will induce the paramagnetic phase, and weaker local fields will result in the antiferromagnetic phase. In the rest of this work, we set the energy constant $J = \hbar = 1$.

The Hamiltonian in Eq.~(\ref{TIM}) can also be written as the sum of nearest neighbor terms    
\begin{equation}
	\Ham = \sum_{l=1}^{L-1} \Ham_l, \label{eq_local_Ham}
.\end{equation} 
Each local Hamiltonian term $\Ham_l$ is 
\begin{equation}\label{Trotter_block}
	\Ham_l = J\sigma_l^z \sigma_{l+1}^z- h_x(n_l\sigma_l^x + n_{l+1}\sigma_{l+1}^x) -  h_z(n_l\sigma_l^z + n_{l+1}\sigma_{l+1}^z)
,\end{equation}
where $n_l=1$ for $l=1$ and $l=L$, and $n_l=1/2$ otherwise. 

\section{Time Evolution of Spin Systems with NNQS}\label{sec_OptApproaches}
In this section, we describe the two main NNQS approaches used to perform the quantum quenches. 
In both approaches, we consider the time evolution from $t=0$ to $t=t_f$ with time interval $dt$. 
The methods described below are independent of the type of neural networks chosen and as such we will test them for both FNNs and RBMs.

\subsection{Time-dependent variational Monte Carlo approach}\label{sec_tVMC} 
The first approach is a variational method that builds on the \SE equation~\cite{carleo2017solving}. The resulting equation of motion for the neural network parameters is 
\begin{align}
    \frac{d\pmb{\theta} (t)}{dt} = -\im \; \pmb{S}^{-1}(t) \pmb{F}(t). \label{eq_evotheta} 
\end{align}
where $\pmb{S}(t)$ is the regularized geometric tensor defined in Eq.~(\ref{eq_GT}).
The elements of the gradient force $\pmb{F}(t)$ are given by $F_k = \langle E_{loc} O^*_k\rangle - \langle E_{loc}\rangle\langle O^*_k\rangle$ where $E_{loc}$ is the local energy. Similarly to Eq.~(\ref{eq_Oloc}), this quantity is evaluated by 
\begin{align}
E_{loc}(\pmb{x}) = \sum_{\pmb{y}} \mathcal{H}_{\pmb{y},\pmb{x}} \frac{\psi_t(\pmb{y})}{\psi_t(\pmb{x})}.   
\end{align} 
It is convenient to rewrite the gradient force as 
\begin{align}\label{eq:gf}
    \pmb{F} = \pmb{X} \pmb{f} 
\end{align}
where the elements of the vector $\pmb{f}$ are given by   
\begin{align}\label{eq:fk}
    f_k = \frac{1}{\sqrt{N_s}}(E_{loc}({\pmb{x}_k}) - \langle E_{loc}(\pmb{x}) \rangle).   
\end{align}
From Eq.~(\ref{eq_evotheta}) one can numerically integrate it using methods like Runge-Kutta. In this work, we use the fourth-order Runge-Kutta method.

\subsection{Projected time-dependent variational Monte Carlo approach}\label{sec_p-tVMC} 
The second approach is fundamentally different~\cite{gutierrez2022real}. In this method, we consider the \SE equation $|\psi_{t'}\rangle = U|\psi_t\rangle$ where $U=e^{-\im \Ham(t'-t)}$ is the unitary evolution operator from time $t \to t^\prime$ and $\Ham$ is the (time-independent) Hamiltonian describing the system dynamics ~\cite{Hamiltonians}.
We can therefore rely on a well-designed cost function $C^U_{\psi_{t},\psi_{t'}}$ to minimize the distance between the known current NNQS $ |\psi_t\rangle$ and the unknown future NNQS $ |\psi_{t'}\rangle$.  
The overlap between the wave functions obtained from the two NNQS is  
\begin{align} \label{eq_overlap}  
	C^U_{\psi_{t},\psi_{t'}} &=\frac{\langle \psi_{t'}|U|\psi_{t}\rangle \langle \psi_{t}|U^{\dag}|\psi_{t'}\rangle}{\langle \psi_{t'}|\psi_{t'} \rangle \langle \psi_{t}|\psi_{t} \rangle} \nonumber \\
	&= \left( \frac{\sum_{\pmb{x},\pmb{x}'}\langle \psi_{t'}|\pmb{x}\rangle \langle \pmb{x}|U|\pmb{x}'\rangle \langle \pmb{x}'|\psi_{t}\rangle}{\sum_{\pmb{z}}|\psi_{t'}(\pmb{z})|^2} \right) \nonumber \\ 
	&\times
	\left( \frac{\sum_{\pmb{y},\pmb{y}'}\langle \psi_{t}|\pmb{y}\rangle \langle \pmb{y}|U^{\dag}|\pmb{y}'\rangle \langle \pmb{y}'|\psi_{t'}\rangle}{\sum_{\pmb{w}}|\psi_t(\pmb{w})|^2} \right) \nonumber \\ 
	&=   \sum_{\pmb{x},\pmb{y}}P_{\psi_{t'}}(\pmb{x})P_{\psi_{t}}(\pmb{y})E^U_{\psi_{t}\psi_{t'}}(\pmb{x})E^U_{\psi_{t'}\psi_{t}}(\pmb{y})  \nonumber \\ 
	&=   \sum_{\pmb{x}} P_{\psi_{t'}}(\pmb{x}) E^U_{loc}(\pmb{x}),
\end{align}
where $E_{\psi_t \psi_{t^\prime}}^U$ and $E_{\psi_{t^\prime} \psi_{t}}^U$ can be interpreted as the local temporal overlaps
\begin{align}
E^U_{\psi_{t}\psi_{t'}}(\pmb{x}) = \sum_{\pmb{x}'}U_{\pmb{x}\pmb{x}'}\psi_{t}(\pmb{x}')/\psi_{t'}(\pmb{x}), \nonumber \\ 
E^U_{\psi_{t'}\psi_{t}}(\pmb{y}) = \sum_{\pmb{y}'}U^{\dag}_{\pmb{y}\pmb{y}'}\psi_{t'}(\pmb{y}')/\psi_{t}(\pmb{y}) 
,\end{align} 
the probabilities are defined from 
\begin{align}
P_{\psi_{t}}(\pmb{x}) = \frac{\langle \psi_{t}|\pmb{x}\rangle \langle \pmb{x}|\psi_{t}\rangle}{\sum_{\pmb{w}}|\psi_t(\pmb{w})|^2}, 
\end{align}
and 
\begin{align}
E^U_{loc}(\pmb{x}) =  \sum_{\pmb{y}}P_{\psi_{t}}(\pmb{y})E^U_{\psi_{t}\psi_{t'}}(\pmb{x})E^U_{\psi_{t'}\psi_{t}}(\pmb{y}).      
\end{align}
For small system sizes, we can consider all $2^L$ possible configurations when calculating the $\pmb{X}$ matrix in Eq.~(\ref{eq_x}). We refer to this explicitly as {\it full summation}. Since all configurations are explored, the gradient always points in the right direction. Given an appropriate learning rate, the overlap $C^U_{\psi_t, \psi_{t^\prime}}$ will converge to $1$ as the number of optimization steps increases. 
When $L$ is large, full summation becomes computationally unfeasible. In these scenarios, Monte Carlo sampling is performed on both NNQS at time $t$ with probabilities $P_{\psi_{t}}$ and $t^\prime = t + dt$ with probabilities $P_{\psi_{t'}}$. In each optimization step, samples are produced stochastically and the gradient is calculated based on incomplete information. 
This is somewhat similar to a ground state search algorithm because one can extract from Eq.~(\ref{eq_overlap}) an effective Hamiltonian 
\begin{align}
    \Ham_{\rm eff} = \frac{U|\psi_{t}\rangle \langle \psi_{t}|U^{\dag}}{ \langle \psi_{t}|\psi_{t} \rangle}.  \label{eq_effH} 
\end{align}
However, there is a major computational difference between evaluating the expectation values for the energy in NNQS~\cite{carleo2017solving} and computing either $E^U_{\psi_{t}\psi_{t'}}$ or $E^U_{\psi_{t'}\psi_{t}}$. This is because $U$ and $\Ham_{\rm{eff}}$ are typically dense but $\Ham$ can be sparse. Hence, given a configuration $\pmb{x}$, one would need to consider all other configurations $\pmb{x}'$. It would thus be impractical to use Eq.~(\ref{eq_overlap}) to evaluate the overlap directly.
We thus decompose the unitary evolution operator $U(dt)$ into local operators spanning over a block of $d \geq 2$ sites, $U_{d,l}(dt/2) = e^{-\im \sum_{l}^{l+d-2}\Ham_{l}dt/2}$, where $\Ham_{l}$ is the two-site local Hamiltonian defined in Eq.~(\ref{eq_local_Ham}). For example, in the case of $d=2$, we can apply the second-order Suzuki-Trotter decomposition~\cite{trotter1959,suzuki1976} to obtain 
\begin{align}
	U(dt) &= U_{2,1}(dt/2)U_{2,2}(dt/2)\dots U_{2,L-1}(dt/2) \nonumber \\ 
	&\times U_{2,L-1}(dt/2)U_{2,L-2}(dt/2) \dots U_{2,1}(dt/2)
.\end{align}

In this way, instead of evaluating directly the evolution for a global operator $U(dt)$, we focus on evolving the neural network sequentially with the local operators $U_{d,l}$. Since each operator $U_{d,l}$ couples one configuration to $2^{d}$ others (instead of $2^L$), the memory required to consider all $\pmb{x^\prime}$ configurations is reduced and the computation can be carried out more efficiently. Unless specified otherwise, we use block sizes $d=6$, which corresponds to 64 configurations.

The time evolution of a wave function represented by NNQS thus becomes an optimization problem
\begin{equation}\label{eq_new_loss_func}
	\min_{\pmb{\theta}}\{I^{U_{d,l}}_{\psi_t,\psi_t'}\} = \min_{\pmb{\theta}}\{1 -C^{U_{d,l}}_{\psi_t,\psi_t'}\} 
\end{equation}
for each Suzuki-Trotter block.
The goal is to minimise the infidelity $I^{U_{d,l}}_{\psi_t,\psi_t'}$.

We use the stochastic reconfiguration~\cite{Sorella1998} optimizer to find the parameters $\pmb{\theta}$ that minimize Eq.~(\ref{eq_new_loss_func}). This generally increases the performance of reaching ground states. In our case, not shown, we find that it performs better than Adam \cite{kingma2014adam}.  
The process of updating the parameters $\pmb{\theta}(m)$ at optimization step $m$ is 
\begin{equation}\label{sr_update}
	\pmb{\theta}(m+1) = \pmb{\theta}(m) - \gamma(m) \pmb{S}^{-1}(m) \pmb{F}^U(m), 
\end{equation}
where $\gamma(m)$ is the learning rate and $\pmb{F}^U$ is the gradient force with elements $F^U_k = \langle E^U_{loc} O^*_k\rangle - \langle E^U_{loc}\rangle\langle O^*_k\rangle$, which can be evaluated using Eqs.~(\ref{eq:gf},\ref{eq:fk}). 
Similarly to $\pmb{F}$, we can write the gradient force $\pmb{F}^U$ as 
\begin{align}
    \pmb{F}^U = \pmb{X} f^U
\end{align} 
where the $N_s$ elements of $f^U$ are    
\begin{align}
    f^U_k = \frac{1}{\sqrt{N_s}}(E^U_{loc}({\pmb{x}_k}) - \langle E^U_{loc}(\pmb{x}) \rangle).   
\end{align} 

In summary, for each block $U_{d,l}$, we perform many optimization steps to minimize the infidelity until it falls below a cut-off $\epsilon$. After that, we move on to the next block. In some scenarios, the cut-off cannot be reached within a predetermined maximum number of steps $M_I$. This happens more often for neural networks that are not expressive enough or when the number of samples is inadequate. 

\subsection{Towards large neural networks: minSR}\label{sec_minsr}
Despite the computational advantage of using the local block evolution operator $U_{d,l}$, the overall optimization routine can still be challenging, especially for larger system sizes.
The main challenge lies in calculating the inverse of $\pmb{S}$. The size of the $\pmb{S}$ matrix and its conditioning determine the computation difficulty in performing the inverse. The latter can be alleviated by the regularization introduced previously. The former, however, is a problem directly related to the sizes of the system and network. The size of $\pmb{S}$ is an $N_p\times N_p$ matrix, where $N_p$ is the number of parameters. This can be very costly for a large network as the time complexity of the inverse operation scales as $\mathcal{O}(N_p^3)$.     
If the number of parameters is larger than the number of samples in the algorithm, i.e. $N_p>N_s$, it is useful to adopt the minSR approach~\cite{chen2023efficient, rende2023simple}. 

For large number of parameters, minSR can significantly reduce the computational costs. For example, in Eq.~(\ref{sr_update})
\begin{align}
(\pmb{S}+\lambda \mathbbm{1}_p)^{-1}\pmb{F}^U &= (\pmb{X}\pmb{X}^{\dag} +\lambda \mathbbm{1}_p)^{-1}\pmb{X}\pmb{f}^U_{E} \nonumber \\
&= \pmb{X}(\pmb{X}^{\dag}\pmb{X} +\lambda \mathbbm{1}_s)^{-1}\pmb{f}^U_{E} \label{eq_minSR} 
\end{align}
where $\mathbbm{1}_p$ is of the size $N_p\times N_p$ and $\mathbbm{1}_s$ is of the $N_s\times N_s$~\cite{rende2023simple}.  
The complexity of evaluating the gradient using Eq.~(\ref{eq_minSR}) is therefore reduced to $\mathcal{O}(N_s^3)$.

\subsection{Towards large neural networks: Kronecker-factored Approximate Curvature (K-FAC)}\label{sec_k-fac}
An alternative method for calculating gradients for large neural networks is K-FAC \cite{martens2015optimizing}. In this method, the matrix $\pmb{S}$ is approximated by much smaller block matrices, where each block corresponds to a subset of all parameters. The parameters are divided into non-overlapping subsets according to the layers of the neural network. If the parameter subsets overlap, the algorithm becomes similar to SLO~\cite{zhang2023ground}. For each optimization at time $t$, we consider and update one non-overlapping subset of parameters $\pmb{\theta}^{k,p,q}$, where all the nodes in layer $k$ are connected to the $k+1$ layer nodes in the range $p \to q$. i.e. $\pmb{\theta}^{1,1,40}$ connects all the nodes in layer 1 to the first 40 nodes in layer 2; $\pmb{\theta}^{2,41,80}$ connects all the nodes in layer 2 to the $41$st to $80$th nodes in layer 3. The new $\pmb{S}'$ matrix for this subset of parameters then becomes $\pmb{S}' = \pmb{S}^{k,p,q}$, which, as mentioned before, is also regularized. The computation of $\pmb{S'}^{-1}$ becomes feasible due to the reduced $N_p$ in each subset.
After updating this subset, we move to the next subset of parameters, until all the parameters have been updated.


\section{Figures of merit} \label{sec_figuresMerit}
We use different indicators to evaluate the performance of our algorithms described in Sec.~\ref{sec_tVMC} and Sec.~\ref{sec_p-tVMC}. 

We integrate the exact infidelity of the trained target state $\psi$, $\mathcal{I}_e$.
The exact infidelity is computed by comparing the NNQS wave function to the exactly time-evolved wave function $\psi_{e,t}$ obtained from exact diagonalization 
\begin{equation}\label{eq_Ie}
	I_e(t) = \Bigg| 1 - \frac{\langle\psi_t|\psi_{e,t}\rangle\langle\psi_{e,t}|\psi_t\rangle}{\langle\psi_t|\psi_t\rangle\langle\psi_{e,t}|\psi_{e,t}\rangle} \Bigg|
\end{equation}
\begin{equation}\label{eq_I_int}
	\mathcal{I}_e(t) = \int_0^t I_e(t) dt
\end{equation}
where $I_e(t)$ represents the infidelity at time $t$. 

We also consider the accumulated errors during optimization Eq.~(\ref{eq_new_loss_func}), computed from the trained target state $\ket{\psi_{t^\prime}}$ and the time-evolved state $U\ket{\psi_{t}}$ 
\begin{equation}\label{eq_R_int}
	\mathcal{R}(t) = \int_0^t \sum_l I^{U_{d,l}}_{\psi_t,\psi_t'} dt
.\end{equation}
Both quantities characterize the deviation from an accurate expression of the wave function over long times. However, the latter uses as a reference a state which may no longer be accurate. 

To understand the errors in the NNQS representation, we calculate the amplitude ratio $A_n(t)$ given by 
\begin{equation}\label{eq_amplitude}
	A_n(t) = \Bigg|\frac{\psi_t(\pmb{x}_M)}{\psi_t(\pmb{y}_n)}\Bigg|
,\end{equation}
and phase distance $D_n(t)$
\begin{align}\label{eq_phase}
    	D_n(t) &= \min(d_n(t),2\pi - d_n(t)), 
\end{align} 
with 
\begin{align}
        d_n(t) &= |\Arg(\psi_t(\pmb{x}_M)) - \Arg(\psi_t(\pmb{y}_n))|, \nonumber  
\end{align}
where $\pmb{x}_M$ and $\pmb{y}_n$ are the most probable and $n$-th most probable configurations obtained from $| \psi_{e,t_f} \rangle$ respectively, and $\Arg$ gives the phase of the probability amplitude.
If the NNQS time evolution is accurate, it should produce the same $A_n(t)$ and $D_n(t)$ as exact diagonalization. 

Lastly, we benchmark the time evolution of a local observable to understand how the error accumulation affects measurable quantities. 
Our choice is to compare the evolution of $\langle \sigma^x_l(t)\rangle$ and/or $\langle \sigma^z_l(t)\rangle$ from our algorithms to results obtained from exact diagonalization or matrix product states. 

\section{Results}\label{sec_results} 
We start with the ground state of the system in the paramagnetic phase $|\psi_{gs}\rangle = |\rightarrow, \rightarrow, \cdots, \rightarrow \rangle$, for a Hamiltonian with only a non-zero and positive $h_x$. The technique used to obtain this ground state in the NNQS expression is detailed in Ref.~\cite{zhang2023ground}.
We then perform a quantum quench from this paramagnetic ground state to a deeply antiferromagnetic system with $h_z = h_x = 0.5 J$.

In the following subsections, we investigate the roles of the time evolution algorithm (i.e. tVMC, p-tVMC), the number of samples $N_s$, and the number of parameters $N_p$.
In this section, we show the quench dynamics obtained from both FNN and RBM. Our goal is not to compare FNN and RBM but to show that our analysis of the hyperparameters applies to generic neural networks.
Lastly, we amalgamate our findings and present the quench dynamics of a large $L=40$ tilted Ising model.

We choose the following default hyperparameters for the FNN such that the network has four layers (one input/visible, two hidden, and one output) with $[L, 4L, 3L, 1]$ nodes in each layer; for the RBM the network has $\alpha L$ hidden nodes where $\alpha = 5$. For both FNN and RBM, the initial learning rate is $\gamma_0 = 0.2$ and it is reduced to $\gamma = 0.8\gamma$ after every 400 optimization steps; the maximum number of optimization steps is $M_I=1000$; the infidelity cut-off is $\epsilon = 10^{-5}$; the time interval is $dt = 0.1$ and each Suzuki-Trotter block $U_{d,l}$ contains size $d=6$ sites.
If any hyperparameter or initial condition is different from the default, it will be explicitly stated in the figure or caption. 
We repeat each simulation 10 times (except for the full summation method and the 40-spin system). The simulations with the highest and the lowest values are subsequently removed. The results are presented as the average of the remaining eight runs, with the error bars indicating the 10th and 90th percentile. 

\subsubsection{Comparison between tVMC and p-tVMC}\label{sec_tVMC_p-tVMC}
\begin{figure}
	\centering
	\includegraphics[width=\columnwidth]{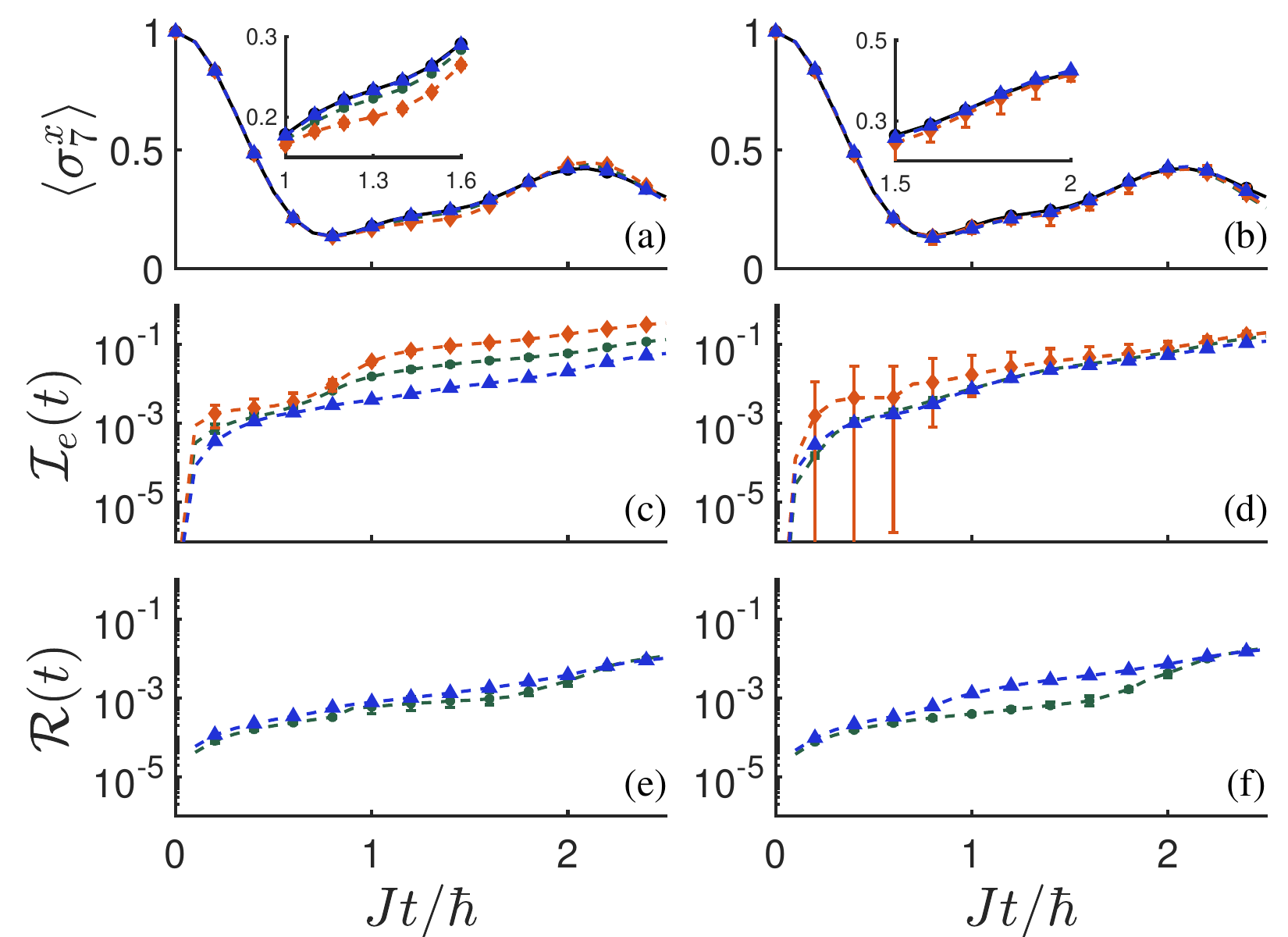}
	\caption{Time evolution of 1D 14-spins tilted Ising model after a quench simulated with different numerical methods. Two architectures of neural networks used are a FNN with layer size $[14, 56, 42, 1]$ and an RBM with $\alpha = 5$, depicted in the left and right columns, respectively. The time interval is $dt = 10^{-1}$ for p-tVMC and $dt = 10^{-3}$ for tVMC.  (a-b) Time evolution of $x$-direction of magnetization of the middle site, $\langle \sigma_7^x\rangle$; (c-d) time evolution of integrated exact infidelity $\mathcal{I}_e(t)$; (e-f) time evolution of accumulated error $\mathcal{R}(t)$. 
    The symbol: 
    \textcolor{black1}{$\bullet$} 
    represents the result obtained from exact diagonalization; 
    \textcolor{red1}{$\blacklozenge$} 
    is from tVMC; 
    \textcolor{green4}{$\ast$} 
    is from p-tVMC and 
    \textcolor{blue4}{$\blacktriangle$} 
    is from full summation. The other hyperparameters are $\epsilon = 10^{-5}$ and $N_s = 10^4$.} \label{fig_fig2}
\end{figure}

\begin{figure*}
	\centering
	\includegraphics[width=2\columnwidth]{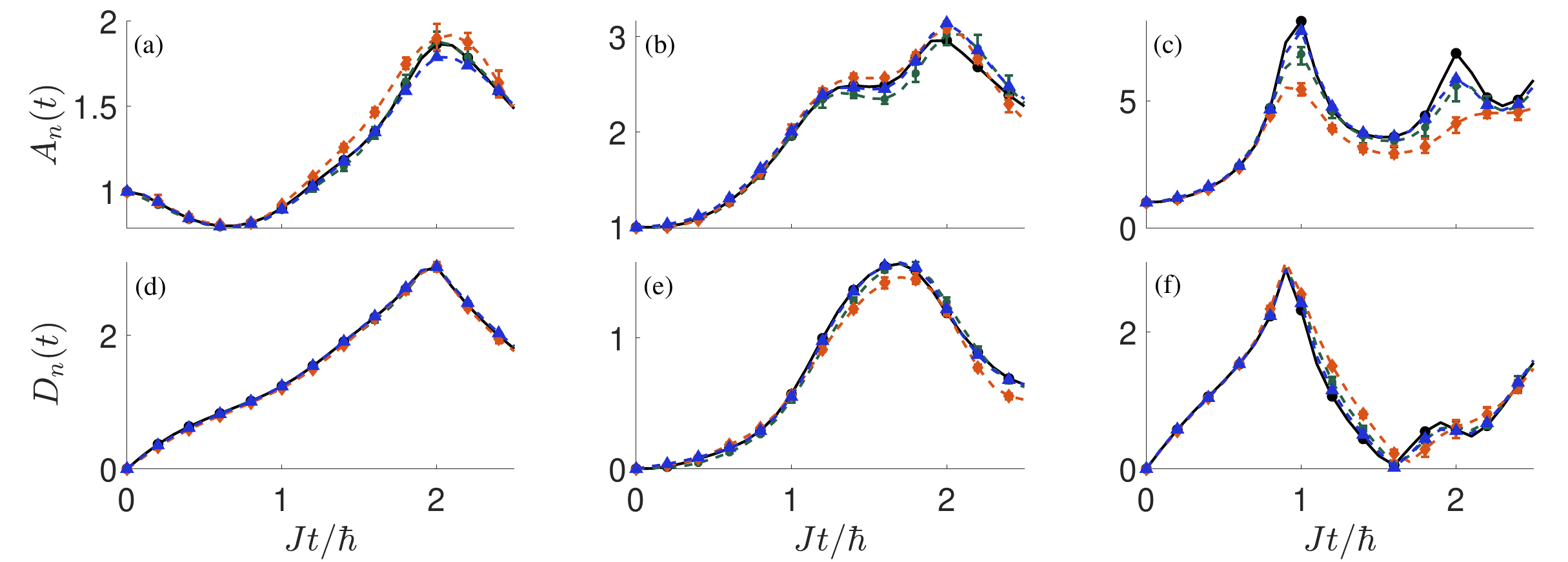}
	\caption{The FNN amplitude ratio $A_n(t)$ and the phase distance $D_n(t)$ of the most probable configuration $\pmb{x}_M$ to the configuration $\pmb{y}_n$ with $n$-th largest probability: (a,d) n = 2, (b,e) $n = 50$, and (c,f) $n = 500$. The symbol \textcolor{blue1}{$\bullet$} represents the result obtained by exact diagonalization, \textcolor{red1}{$\blacklozenge$} is for tVMC, \textcolor{green1}{$\ast$} and \textcolor{purple1}{$\blacktriangle$} indicate the results of p-tVMC by Monte Carlo method and the full summation, respectively.} \label{fig_fig2_2}
\end{figure*}

We first compare in Fig.~\ref{fig_fig2} the quench dynamics obtained from tVMC (Sec.~\ref{sec_tVMC}) and p-tVMC (Sec.~\ref{sec_p-tVMC}). The system size is $L = 14$ so we also have the option to perform full summation. 
The left column (a, c, e) contains results from FNN and the right column (b, d, f) contains results from RBM.

In panels (a,b), we plot the expectation value of a mid-chain local observable $\langle \sigma^x_7 \rangle$. All the NNQS methods predict the same trends in the dynamics of $\langle \sigma^x_7 \rangle$. However, closer comparisons with the results from exact diagonalization, shown in the insets, confirm that p-tVMC performs better than the tVMC.
In panels (c,d), we show the integrated exact infidelity $\mathcal{I}_e$ from Eq.~(\ref{eq_I_int}). The infidelity of p-tVMC is always lower than that of tVMC.
This means that the p-tVMC wave function has a higher overlap with the exact diagonalization wave function. Having a lower $\mathcal{I}_e$ means that the NNQS wave function will continue to produce correct local observables for a longer time.
The full summation method further removes all Monte Carlo errors and therefore performs the best. 
We also point out that p-tVMC demonstrates greater stability compared to tVMC. For tVMC to produce accurate results, a small time interval $dt = 0.001$ needs to be used. 
This need arises in tVMC due to the lack of an optimization process. As a result, errors tend to build up more quickly and the simulation may fail when $dt$ is larger.
Another reason contributing to the instability of the simulation is the choice of the regularization parameter $\lambda$ for $\pmb{S}$ matrix: when $\lambda$ is large, the simulation is stable but inaccurate. The inaccuracy is sizeable already for $\lambda = 10^{-6}$; when $\lambda$ is appropriately chosen, the simulation is both stable and accurate. However, if $\lambda$ is too small (e.g., $\lambda \approx 10^{-8}$), the simulation is unstable so that jump instabilities occur over time ~\cite{hofmann2022role}. For these reasons, here we do not perform this type of regularization, i.e., we set $\lambda = 0$ ~\cite{regularization} for tVMC.
In panels (e, f), we show the accumulated error $\mathcal{R}$ from Eq.~(\ref{eq_R_int}) as a function of $t$.
Unlike $\mathcal{I}_e$ which compares the quality of the NNQS wave function to the exact diagonalization wave function, $\mathcal{R}$ indicates how well the NNQS wave function $| \psi_t \rangle$ overlaps with the time-evolved version of itself  $| \psi_t^\prime \rangle = U| \psi_t \rangle$. 
An increase in $\mathcal{R}$ similarly conveys that the NNQS wave function is unable to converge to a good target state. Therefore $\mathcal{R}$ and $\mathcal{I}_e$, although different in magnitude, can be used interchangeably to give insights into the accuracy of the predicted quench dynamics. If the system is large and calculating $|\psi_e \rangle$ and $\mathcal{I}_e$ becomes unfeasible, we can still rely on $\mathcal{R}$ to gauge the quality of the NNQS wave function. 

To understand the errors in the tVMC, p-tVMC, and full summation methods, we plot the FNN evolution of probability amplitude ratio $A_n(t)$ and phase distance $D_n(t)$ in Fig.~\ref{fig_fig2_2}.
In panels (a, d), where $n = 2$, we find that all NNQS methods are capable of producing $A_2(t)$ and $D_2(t)$ that are similar to exact diagonalization, but the results from p-tVMC perform better than tVMC.
In panels (b, e), where $n = 50$, the $A_{50}(t)$ and $D_{50}(t)$ produced from NNQS methods begin to show different levels of deviation from exact diagonalization.
Comparing tVMC, p-tVMC, and full summation, the last two start to deviate at a later time. This is more obvious from $D_{50}(t)$ in panel (e).
In panels (c, f), where $n = 500$, the degree of deviation varies largely among tVMC, p-tVMC, and full summation.
In general, p-tVMC and full summation are capable of producing more accurate probability amplitudes.
One may argue that the correct prediction of these states is less important due to their diminishing probability amplitude.
However, in quench dynamics, unlikely configurations can become more likely over time.
The inability to produce accurate probability amplitudes for less probable states can have major long-term consequences.

\subsubsection{Comparison among number of samples}\label{sec_comp_samples}
\begin{figure}
	\includegraphics[width=\columnwidth]{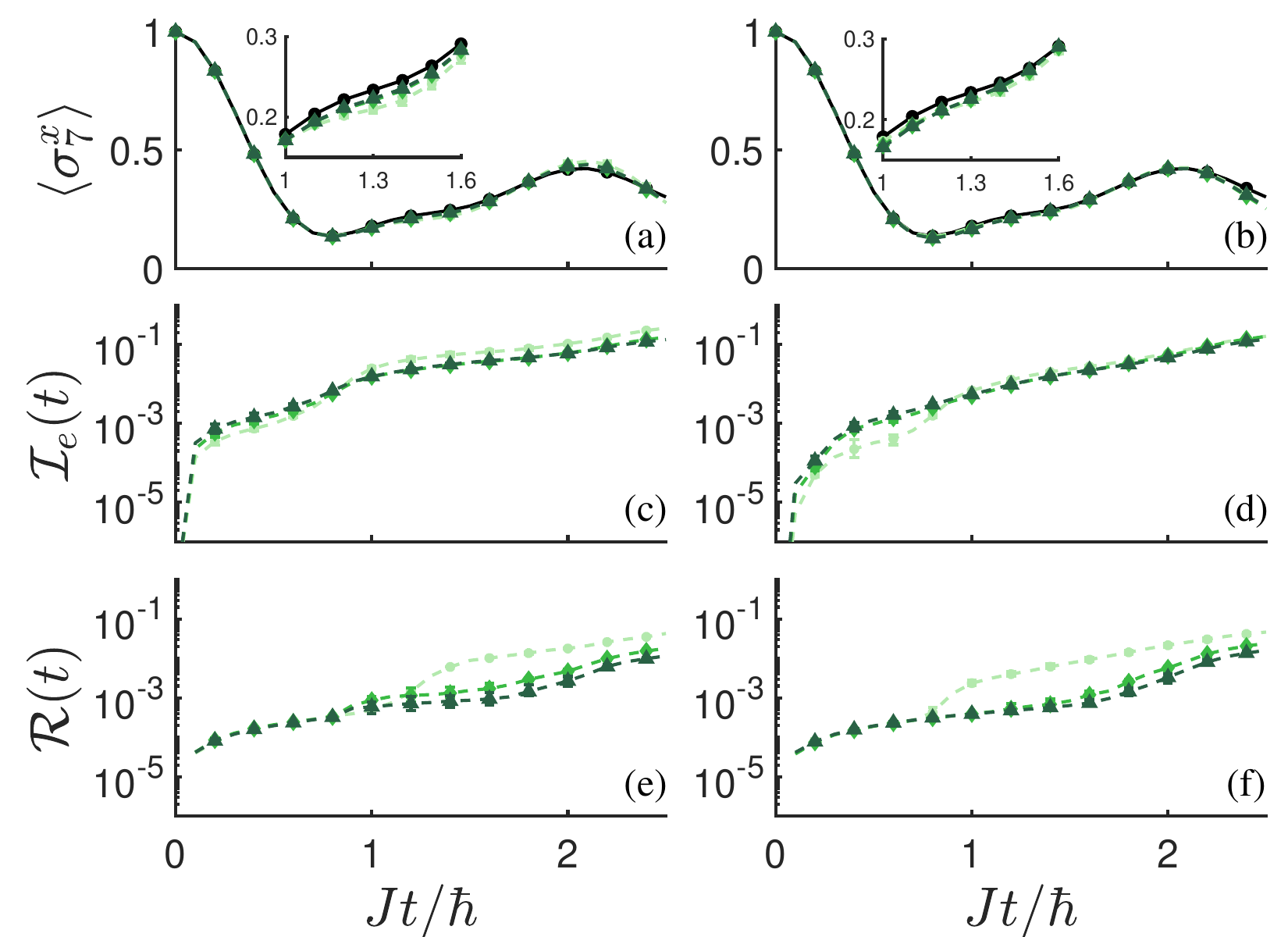}
	\caption{p-tVMC time evolution of 1D 14-spins tilted Ising model after quenching with different numbers of samples. Two architectures of neural networks are considered, FNN with layer size $[14,56,42, 1]$ and RBM with $\alpha = 7$, depicted in the left and columns, respectively. (a-b) Time evolution of $x$-direction of magnetization of the middle site $\langle \sigma_7^x\rangle$; (c-d) time evolution of integrated exact infidelity $\mathcal{I}_e(t)$; (e-f) time evolution of accumulated error $\mathcal{R}(t)$. The symbol  
    \textcolor{black1}{$\bullet$}
    represents the exact value, while  
    \textcolor{green2}{$\ast$}, \textcolor{green3}{$\blacklozenge$} and \textcolor{green4}{$\blacktriangle$} 
    corresponds to the different number of samples: $N_s = \{10^3, 5\times10^3, 10^4\}$. } \label{fig_fig3}
\end{figure}

\begin{figure*}
    \includegraphics[width=2\columnwidth]{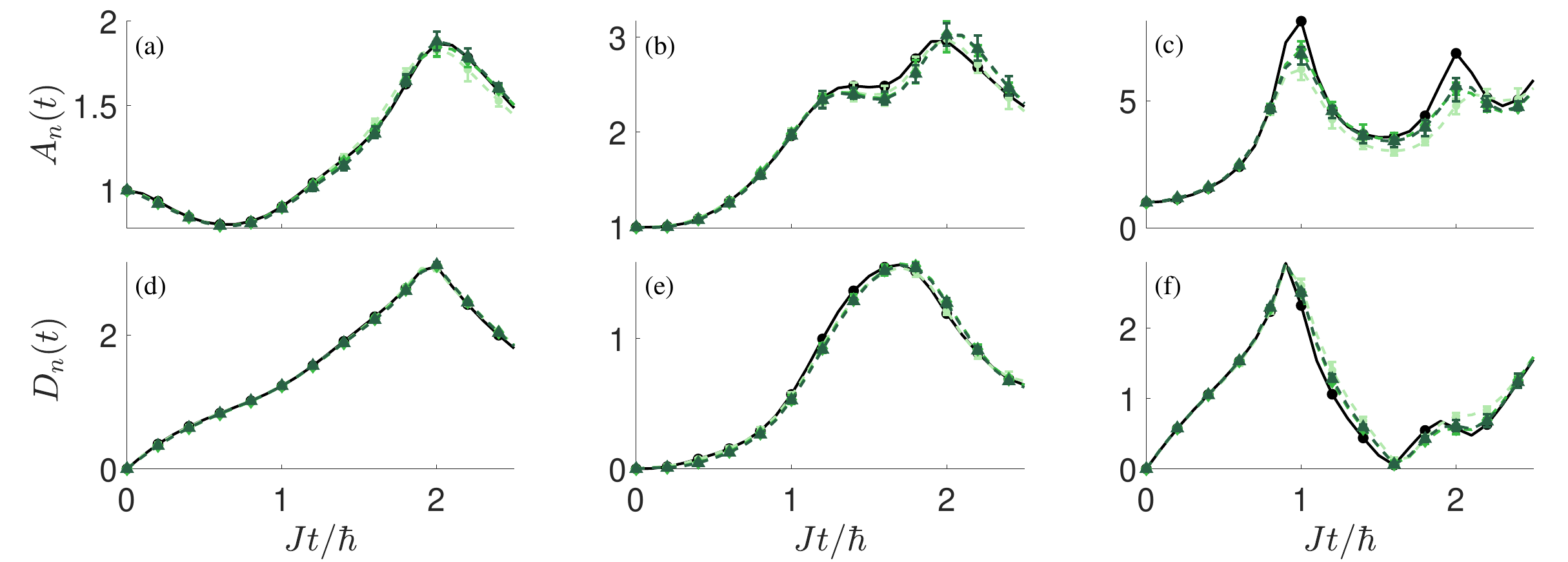}
	\caption{The FNN amplitude ratio $A_n(t)$ and the phase distance $D_n(t)$ of the most probable configuration $\pmb{x}_M$ to the configuration $\pmb{y}_n$ with $n$-th largest probability: (a,d) n = 2, (b,e) $n = 50$, and (c,f) $n = 500$.
    \textcolor{black1}{$\bullet$}
    represents the exact value,  
    \textcolor{green2}{$\ast$}, \textcolor{green3}{$\blacklozenge$} and \textcolor{green4}{$\blacktriangle$}  
    indicate the number of samples: $N_s = \{10^3, 5\times10^3, 10^4\}$. } \label{fig_fig3_2}
\end{figure*}

We investigate the effects of varying the numbers of samples in Fig. \ref{fig_fig3}. All NNQS simulations are done with p-tVMC and the system size is $L = 14$. The left column (a, c, e) contains results from FNN and the right column (b, d, f) contains results from RBM. 

In panels (a,b), we plot the expectation value of a mid-chain local observable $\langle \sigma^x_7 \rangle$. 
All the lines are close to the exact value, although we can see
a larger improvement as the sample size $N_s$ increases from $10^3$ to $5\times 10^3$. 
In panels (c,d), we show the integrated exact infidelity $\mathcal{I}_e$ Eq.~(\ref{eq_I_int}). The exact infidelity improves more noticeably from $N_s = 10^3$ to $5\times 10^3$ but further increasing to $N_s = 10^4$ has no further benefits.
This is not surprising because Monte Carlo sampling error typically scales with $\sqrt{1/N_s}$. 
In panels (e, f), we show the accumulated error $\mathcal{R}$ Eq.~(\ref{eq_R_int}) as a function of $t$. We notice a jump in $\mathcal{R}(t)$ for $N_s = 10^3$ which occurs around $t=1.2$ in FNN and $t=0.8$ in RBM. As $N_s$ increases, the jump is delayed to a larger $t$ and reduced in magnitude. Similarly, we do not observe notable improvements between $N_s = 5\times 10^3$ and $10^4$. 

We plot the FNN evolution of $A_n(t)$ and $D_n(t)$ in Fig.~\ref{fig_fig3_2}.
In panels (a, d), where $n = 2$, the predicted probability amplitude is good for all numbers of samples.
In panels (b, e), where $n = 50$, the errors for $N_s = 10^3, 5\times10^3$, and $10^4$ are still small. Although the lines are not exactly on top of each other, their error bars overlap and it is difficult to tell them apart statistically.
In panels (c, f), where $n = 500$, the predicted probability amplitudes of this less probable configuration $A_{500}$ are not satisfactory for all $N_s = 10^3, 5\times10^3, 10^4$.
The accuracy in $A_{500}$ and $D_{500}$ improves only slightly as $N_s$ increases from $N_s=10^3$ to $N_s=5\times10^3$.
We should point out that the dependence on $N_s$ examined in Figs.~\ref{fig_fig3},\ref{fig_fig3_2} is also affected by the fact that we are here considering a small system size $L$ (which was chosen so that we can also consider the full summation case for our investigations). Although the Metropolis sampling in p-tVMC can explore the full configuration space after repeated optimization steps, the samples produced in each optimization step within each unitary block do not cover the full configuration space. Therefore, the update of parameters $\pmb{\theta}$ is still based on incomplete information in p-tVMC.

\subsubsection{Comparing different numbers of neural network parameters}
\begin{figure}
	\centering
	\includegraphics[width=\columnwidth]{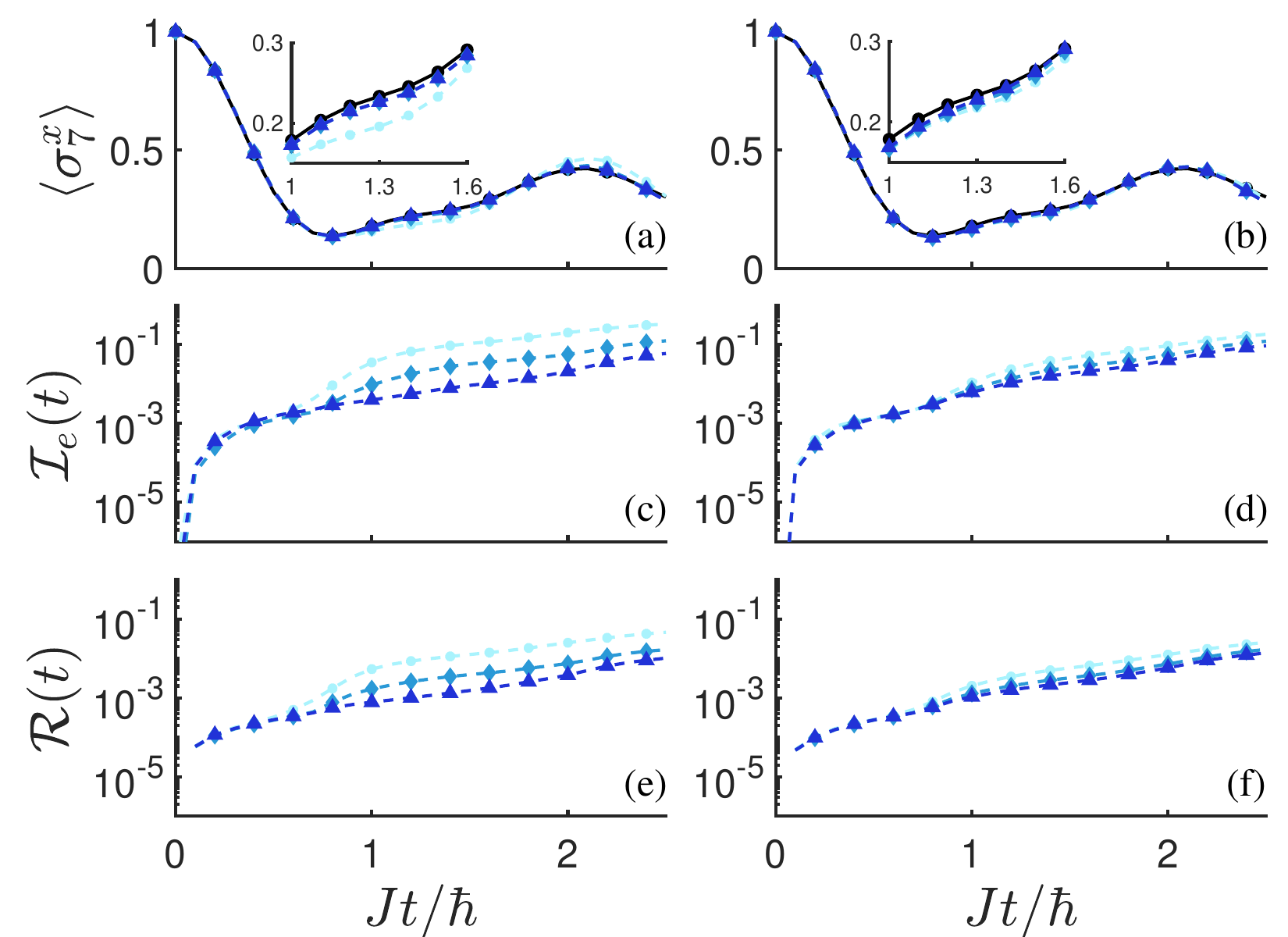}
	\caption{Full summation time evolution of 1D 14-spins tilted Ising model after quenching with the different number of parameters. Two architectures of neural networks are employed, FNN and RBM depicted in the left and columns, respectively. (a-b) Time evolution of $x$-direction of magnetization of the middle site $\langle \sigma_7^x\rangle$; (c-d) time evolution of integrated exact infidelity $\mathcal{I}_e(t)$; (e-f) time evolution of accumulated error $\mathcal{R}(t)$. 
    The symbol  
    \textcolor{black1}{$\bullet$} 
    represents the result obtained from exact diagonalization;   
    \textcolor{blue2}{$\ast$} 
    is for the layer size $[14,14,14,1]$ with $N_p = 435$ in FNN and $\alpha = 3$ in RBM with $N_p = 644$; 
    \textcolor{blue3}{$\blacklozenge$}
    represents the layer size $[14,42,28,1]$ with $N_p = 1863$ in FNN and $\alpha = 5$ in RBM with $N_p = 1064$; 
    \textcolor{blue4}{$\blacklozenge$}
    for the layer size $[14,56,42,1]$ with $N_p = 3277$ in FNN and $\alpha = 7$ in RBM with $N_p = 1484$. Here, we use the full summation over the $N_s = 2^{14}$ configurations. }\label{fig_fig4}
\end{figure}

\begin{figure*}
	\includegraphics[width=2\columnwidth]{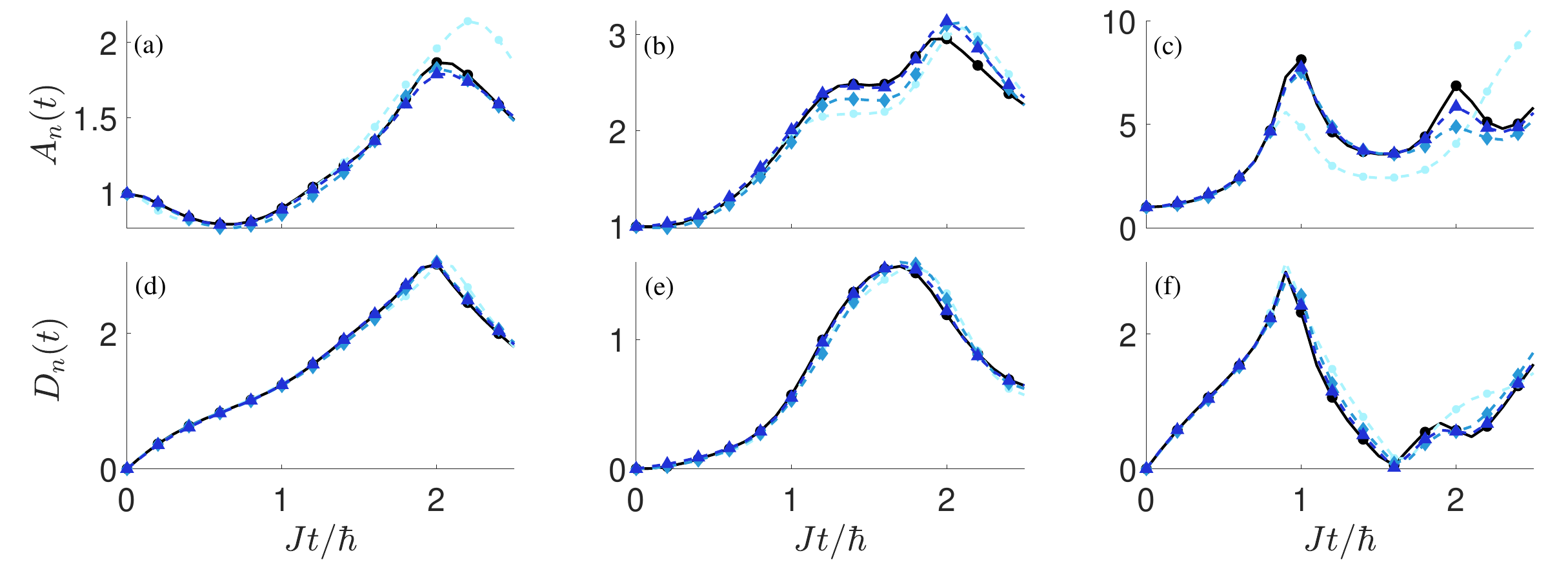}
	\caption{The FNN amplitude ratio $A_n(t)$ and the phase distance $D_n(t)$ of the most probable configuration $\pmb{x}_M$ to the configuration $\pmb{y}_n$ with $n$-th largest probability: (a,d) n = 2, (b,e) $n = 50$, and (c,f) $n = 500$. 
    The symbol  
    \textcolor{black1}{$\bullet$} 
    represents the result obtained from exact diagonalization; 
    \textcolor{blue2}{$\ast$} 
    is for the layer size: $[14,14,14,1]$ with $N_p = 435$ in FNN and $\alpha = 3$ in RBM with $N_p = 644$; 
    \textcolor{blue3}{$\blacklozenge$}
    represents the layer size: $[14,42,28,1]$ with $N_p = 1863$ in FNN and $\alpha = 5$ in RBM with $N_p = 1064$; 
    \textcolor{blue4}{$\blacklozenge$}
    for the layer size: $[14,56,42,1]$ with $N_p = 3277$ in FNN and $\alpha = 7$ in RBM with $N_p = 1484$. Here, we use the full summation over the $N_s = 2^{14}$ configurations.
    } \label{fig_fig4_2}
\end{figure*}

In Fig.~\ref{fig_fig4}, we vary the number of neural network parameters $N_p$ and study the effects it has on the quench dynamics. The left column (a, c, e) contains results from FNN and the right column (b, d, f) contains results from RBM. In all panels, we use the full summation method with all $N_s = 2^{14}$ unique configurations to evolve the NNQS so that the results are independent of the Monte Carlo sampling noise.

In panels (a,b), we plot the expectation value of a mid-chain local observable $\langle \sigma^x_7 \rangle$. In the inset, results from the FNN with the lowest $N_p$ deviate the most from the results obtained via exact diagonalization. The simulation of the quench dynamics becomes more accurate as $N_p$ increases.
In panels (c,d), we show the integrated exact infidelity $\mathcal{I}_e$ Eq.~(\ref{eq_I_int}).
We see more clearly that increasing $N_p$ can effectively reduce the $\mathcal{I}_e$.
In FNN, increasing $N_p$ from $435$ to $1064$ improves $\mathcal{I}_e$ by an order of magnitude.
The benefit of increasing $N_p$, at least for the system sizes considered, is far more evident than $N_s$. We also relate this to the fact that in p-tVMC multiple optimizations are performed, and in each one of them new samples are produced.
In panels (e,f), we present the accumulated error $\mathcal{R}$ Eq.~(\ref{eq_R_int}).
Similarly, $\mathcal{R}$ is reduced as $N_p$ increases.
The successful suppression of $\mathcal{I}_e$ and $\mathcal{R}$ suggest that accurate time evolution could be achieved over a longer time.
Interestingly, although increasing $N_p$ has similar benefits in RBM, the magnitude of improvement is less.

We plot the FNN evolution of $A_n(t)$ and $D_n(t)$ in Fig.~\ref{fig_fig4_2}.  
In panels (a, d), where $n = 2$, the smallest neural network lacks the expressibility to correctly represent even the most probable configurations.
As $N_p$ increases, $A_{20}$ 
follows closely the exact diagonalization results.
In panels (b, e), where $n = 50$, we observe a deviation from the exact diagonalization results for all $N_p$ used but higher $N_p$ deviates from the exact diagonalization results at later $t$.
These deviations may arise from the increased complexity of the wave function representation. As the system undergoes dynamic evolution, the rapid changes in the wave function pose additional difficulties for the optimization process, resulting in deviations from expected behaviors.
In panels (c, f), where $n = 500$, larger neural networks continue to produce more accurate results. 

In general, the network's expressive capability increases with $N_p$ and becomes more effective in representing the complex correlations in the wave function. However, there is a significant computational downside. In Eq.~(\ref{eq_GT}), $\pmb{S}$ is a $N_p \times N_p$ matrix and the inversion of $\pmb{S}$ requires a time complexity of $\mathcal{O}(N_p^3)$. In Sec.~\ref{sec_minsr}, we have highlighted that minSR can address this computational difficulty as it allows to reshape the $\pmb{S}$ into a $N_s \times N_s$ matrix. 
Furthermore, in Sec.~\ref{sec_comp_samples}, we have shown that the time evolution accuracy is not significantly impacted by $N_s$ beyond an intermediate $N_s$ (although for a small system).
As such, we can rely on minSR to increase $N_p$ while not significantly increasing $N_s$.
The result will be a much more expressive neural network with no significant increase in computational complexity.
An alternative approach to address the computational difficulty is the K-FAC method in Sec.~\ref{sec_k-fac}, following which we can optimize, at each time, only a portion of the neural network.

\begin{figure}
    \includegraphics[width=\columnwidth]{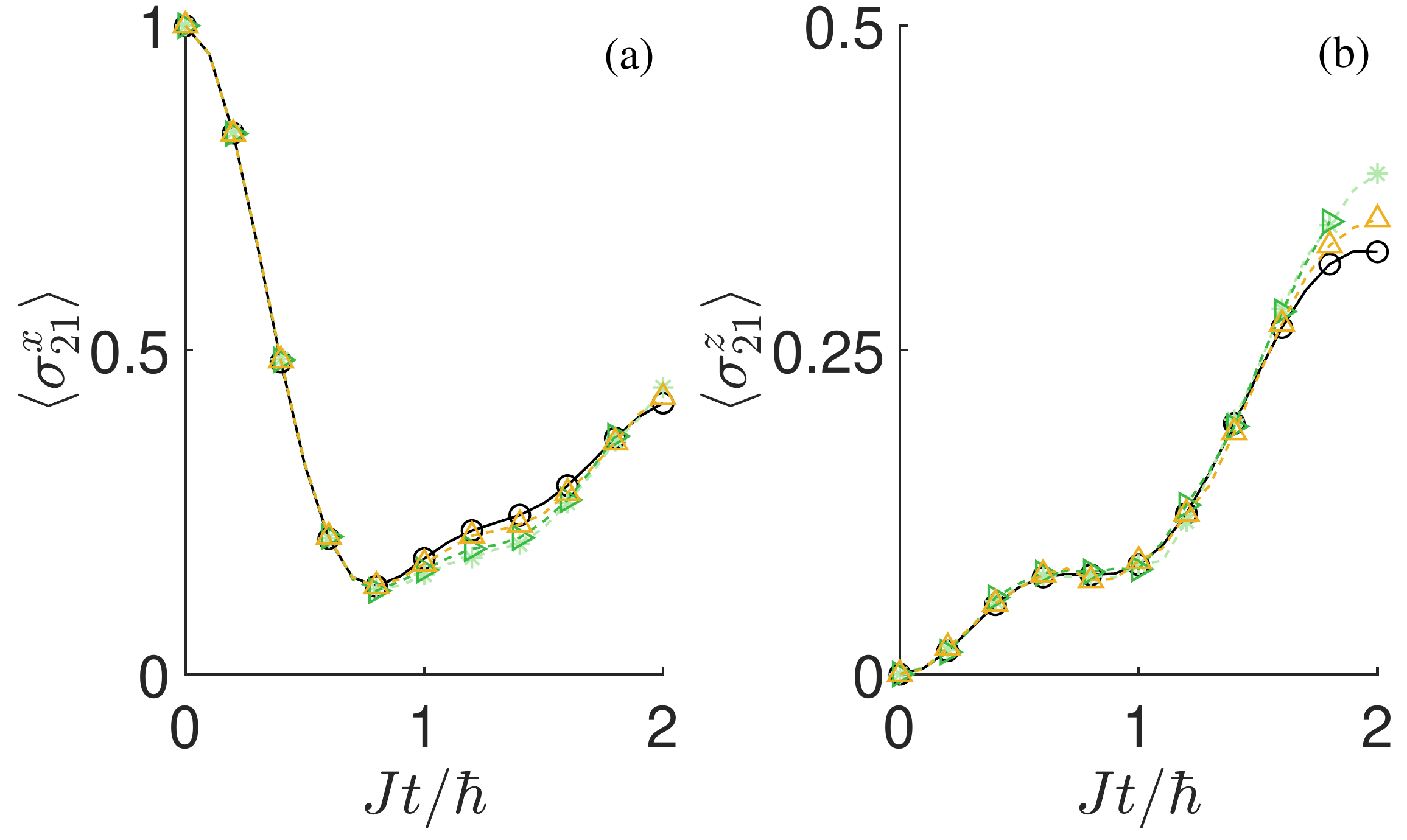}
    \caption{Time evolution of local observable for 40-spin 1D tilted Ising model after quenching with the different optimizers. The FNN architecture is employed. (a) Time evolution of $x$-direction of magnetization of the middle site $\langle \sigma_{21}^x\rangle$; (b) time evolution of $z$-direction of magnetization of the middle site $\langle \sigma_{21}^z\rangle$. The symbol 
    \textcolor{black1}{$\circ$} 
    represents the result obtained from tMPS; 
    \textcolor{green2}{$\ast$}
    represents the K-FAC method with $N_s = 10^4$; 
    \textcolor{green3}{$\triangleright$}
    is for the K-FAC method with $N_s = 10^5$ and with $t_f =1.8$; 
    \textcolor{yellow1}{$\triangle$}
    is for the minSR with $N_s = 10^4$. The parameters are $\epsilon = 5 \times 10^{-6}$ and the layer size [L, 4L, 3L, 1] are used. } \label{fig_40spins}
\end{figure} 

\subsubsection{Larger system size with minSR and K-FAC}
We combine all our insights from the previous sections and apply suitable hyperparameters to investigate the quench dynamics of a $L=40$ tilted Ising model.
For a FNN structure of $[L, 4L, 3L, 1]$, with $N_p \approx 2.6 \times 10^4$, the direct calculation of $\pmb{S}^{-1}$ exceeds computational capacity even on an A100 40GB GPU.
The numerical simulation of this system with large $N_p$ in a single GPU is only possible with K-FAC or minSR. 

In K-FAC, we divide the neural network parameters into $5$ smaller blocks of non-overlapping subsets, and each subset of parameters is updated sequentially during the optimization step. The parameters in the first and the third layers are each treated as a single block. Due to the large number of parameters in the second layer, they are divided into 3 blocks, with each block including these parameters related to 40 hidden nodes. The parameter blocks therefore are $\pmb{\theta} = \{\pmb{\theta}^{1,1,160}, \pmb{\theta}^{2,1,40}, \pmb{\theta}^{2,41,80}, \pmb{\theta}^{2,81,120}, \pmb{\theta}^{3,1,1}\}$. 
In minSR, the size of $\pmb{S}$ becomes $N_s \times N_s$ where $N_s = 10^4$ and this ensures the feasibility of computing the inverse.

In Fig.~\ref{fig_40spins}, we show the FNN results from p-tVMC with either K-FAC or minSR.
We plot the expectation values of mid-chain local observables $\langle \sigma^x_{21} \rangle$ and $\langle \sigma^z_{21} \rangle$ in panels (a) and (b) respectively.
Results from p-tVMC with K-FAC are very similar to the tMPS benchmark up to $t \sim 1$.
For $\langle \sigma^x_{21} \rangle$ beyond $t = 1$, the results from K-FAC start to deviate from the tMPS benchmark.
Comparing the results from K-FAC with $N_s = 10^4$ and $N_s = 10^5$, we see an improvement in the accuracy of $\langle \sigma^x_{21} \rangle$. However, this comes at the expense of doing $9$ more times Monte Carlo sampling. This is in line with our findings from Sec.~\ref{sec_comp_samples}.
The results from p-tVMC with minSR, on the other hand, perform much better. 
For $\langle \sigma^z_{21} \rangle$, the results from K-FAC deviate from the tMPS benchmark at a later time $t = 1.8$. The p-tVMC with minSR approach continues to produce more accurate results.
Using the same number of $N_p$ and $N_s=10^4$, the minSR results are very similar to the tMPS benchmark. We relate this to the fact that with minSR one is optimizing using the full information in the geometric tensor $\pmb{S}$, although it is represented more efficiently thanks to the particular mathematical structure of the problem.

\section{Conclusions}\label{sec_conclusions}   
We have investigated the quench from the paramagnetic to the antiferromagnetic phase of a one-dimension spin $1/2$ tilted Ising model. To evaluate the quality of the predicted wave-function we consider the value of the local magnetization, the infidelity with the exact evolution, and the error from the expected evolution. We also compare the predicted amplitude and phase from the neural network with the exact one for the most probable configurations. 
To increase the general character of the results we consider two different neural networks, a restricted Boltzmann machine and a feed-forward neural network. We show in both cases that p-tVMC allows the use of much larger time steps and is generally more stable. We then focus on p-tVMC and show that both increasing the number of samples and of parameters can significantly help the performance. 
In particular, we show that minSR and K-FAC can both be applied to p-tVMC, allowing stochastic reconfiguration for systems that may require a larger number of parameters.
Although our preliminary results with K-FAC performed worse than minSR, there is potential to improve the K-FAC results further, either by further fine-tuning of its hyperparameters, e.g. different ways to divide the parameters into blocks, or by considering the SLO approach based on dividing the neural network parameters into blocks that overlap with each other~\cite{zhang2023ground}.   
We also highlight that the number of samples required could also be reduced by using control variate~\cite{SinibaldiVicentini2023}.    
Explorations in these directions are intended for future works.

\vspace{1cm}
\section{Acknowledgements}\label{sec_acknowledgements} 
We acknowledge the fruitful discussion with I. Arad, F. Vicentini, L. Wang and Z. Wu. 
D.~P. acknowledges support from the Ministry of Education Singapore, under Grant No. MOE-T2EP50120-0019. 
X.~Xu acknowledges support from joint Israel-Singapore NRF-ISF Research Grant No. NRF2020-NRF-ISF004-3528. 
B.~Xing acknowledges support from the Quantum Engineering Programme NRF2021-QEP2-02-P03. 
W.~Zhang acknowledges support from the National Research Foundation, Singapore, and A*STAR under its CQT Bridging Grant. 
The computational work was partially performed at the National Supercomputing Centre, Singapore~\cite{nscc}.

\bibliography{Bibliography.bib}

\begin{thebibliography}{95}%
\makeatletter
\providecommand \@ifxundefined [1]{%
 \@ifx{#1\undefined}
}%
\providecommand \@ifnum [1]{%
 \ifnum #1\expandafter \@firstoftwo
 \else \expandafter \@secondoftwo
 \fi
}%
\providecommand \@ifx [1]{%
 \ifx #1\expandafter \@firstoftwo
 \else \expandafter \@secondoftwo
 \fi
}%
\providecommand \natexlab [1]{#1}%
\providecommand \enquote  [1]{``#1''}%
\providecommand \bibnamefont  [1]{#1}%
\providecommand \bibfnamefont [1]{#1}%
\providecommand \citenamefont [1]{#1}%
\providecommand \href@noop [0]{\@secondoftwo}%
\providecommand \href [0]{\begingroup \@sanitize@url \@href}%
\providecommand \@href[1]{\@@startlink{#1}\@@href}%
\providecommand \@@href[1]{\endgroup#1\@@endlink}%
\providecommand \@sanitize@url [0]{\catcode `\\12\catcode `\$12\catcode
  `\&12\catcode `\#12\catcode `\^12\catcode `\_12\catcode `\%12\relax}%
\providecommand \@@startlink[1]{}%
\providecommand \@@endlink[0]{}%
\providecommand \url  [0]{\begingroup\@sanitize@url \@url }%
\providecommand \@url [1]{\endgroup\@href {#1}{\urlprefix }}%
\providecommand \urlprefix  [0]{URL }%
\providecommand \Eprint [0]{\href }%
\providecommand \doibase [0]{https://doi.org/}%
\providecommand \selectlanguage [0]{\@gobble}%
\providecommand \bibinfo  [0]{\@secondoftwo}%
\providecommand \bibfield  [0]{\@secondoftwo}%
\providecommand \translation [1]{[#1]}%
\providecommand \BibitemOpen [0]{}%
\providecommand \bibitemStop [0]{}%
\providecommand \bibitemNoStop [0]{.\EOS\space}%
\providecommand \EOS [0]{\spacefactor3000\relax}%
\providecommand \BibitemShut  [1]{\csname bibitem#1\endcsname}%
\let\auto@bib@innerbib\@empty
\bibitem [{\citenamefont {Carleo}\ and\ \citenamefont
  {Troyer}(2017)}]{carleo2017solving}%
  \BibitemOpen
  \bibfield  {author} {\bibinfo {author} {\bibfnamefont {G.}~\bibnamefont
  {Carleo}}\ and\ \bibinfo {author} {\bibfnamefont {M.}~\bibnamefont
  {Troyer}},\ }\bibfield  {title} {\bibinfo {title} {Solving the quantum
  many-body problem with artificial neural networks},\ }\href
  {https://doi.org/10.1126/science.aag2302} {\bibfield  {journal} {\bibinfo
  {journal} {Science}\ }\textbf {\bibinfo {volume} {355}},\ \bibinfo {pages}
  {602} (\bibinfo {year} {2017})}\BibitemShut {NoStop}%
\bibitem [{\citenamefont {Choo}\ \emph {et~al.}(2019)\citenamefont {Choo},
  \citenamefont {Neupert},\ and\ \citenamefont {Carleo}}]{choo2019two}%
  \BibitemOpen
  \bibfield  {author} {\bibinfo {author} {\bibfnamefont {K.}~\bibnamefont
  {Choo}}, \bibinfo {author} {\bibfnamefont {T.}~\bibnamefont {Neupert}},\ and\
  \bibinfo {author} {\bibfnamefont {G.}~\bibnamefont {Carleo}},\ }\bibfield
  {title} {\bibinfo {title} {Two-dimensional frustrated
  ${J}_{1}\text{\ensuremath{-}}{J}_{2}$ model studied with neural network
  quantum states},\ }\href {https://doi.org/10.1103/PhysRevB.100.125124}
  {\bibfield  {journal} {\bibinfo  {journal} {Phys. Rev. B}\ }\textbf {\bibinfo
  {volume} {100}},\ \bibinfo {pages} {125124} (\bibinfo {year}
  {2019})}\BibitemShut {NoStop}%
\bibitem [{\citenamefont {Choo}\ \emph {et~al.}(2020)\citenamefont {Choo},
  \citenamefont {Mezzacapo},\ and\ \citenamefont {Carleo}}]{choo2020fermionic}%
  \BibitemOpen
  \bibfield  {author} {\bibinfo {author} {\bibfnamefont {K.}~\bibnamefont
  {Choo}}, \bibinfo {author} {\bibfnamefont {A.}~\bibnamefont {Mezzacapo}},\
  and\ \bibinfo {author} {\bibfnamefont {G.}~\bibnamefont {Carleo}},\
  }\bibfield  {title} {\bibinfo {title} {Fermionic neural-network states for
  ab-initio electronic structure},\ }\href
  {https://doi.org/10.1038/s41467-020-15724-9} {\bibfield  {journal} {\bibinfo
  {journal} {Nature communications}\ }\textbf {\bibinfo {volume} {11}},\
  \bibinfo {pages} {2368} (\bibinfo {year} {2020})}\BibitemShut {NoStop}%
\bibitem [{\citenamefont {Hermann}\ \emph {et~al.}(2020)\citenamefont
  {Hermann}, \citenamefont {Sch{\"a}tzle},\ and\ \citenamefont
  {No{\'e}}}]{hermann2020deep}%
  \BibitemOpen
  \bibfield  {author} {\bibinfo {author} {\bibfnamefont {J.}~\bibnamefont
  {Hermann}}, \bibinfo {author} {\bibfnamefont {Z.}~\bibnamefont
  {Sch{\"a}tzle}},\ and\ \bibinfo {author} {\bibfnamefont {F.}~\bibnamefont
  {No{\'e}}},\ }\bibfield  {title} {\bibinfo {title} {Deep-neural-network
  solution of the electronic schr{\"o}dinger equation},\ }\href
  {https://doi.org/10.1038/s41557-020-0544-y} {\bibfield  {journal} {\bibinfo
  {journal} {Nature Chemistry}\ }\textbf {\bibinfo {volume} {12}},\ \bibinfo
  {pages} {891} (\bibinfo {year} {2020})}\BibitemShut {NoStop}%
\bibitem [{\citenamefont {Guo}\ and\ \citenamefont
  {Poletti}(2021)}]{guo2021scheme}%
  \BibitemOpen
  \bibfield  {author} {\bibinfo {author} {\bibfnamefont {C.}~\bibnamefont
  {Guo}}\ and\ \bibinfo {author} {\bibfnamefont {D.}~\bibnamefont {Poletti}},\
  }\bibfield  {title} {\bibinfo {title} {Scheme for automatic differentiation
  of complex loss functions with applications in quantum physics},\ }\href
  {https://doi.org/10.1103/PhysRevE.103.013309} {\bibfield  {journal} {\bibinfo
   {journal} {Phys. Rev. E}\ }\textbf {\bibinfo {volume} {103}},\ \bibinfo
  {pages} {013309} (\bibinfo {year} {2021})}\BibitemShut {NoStop}%
\bibitem [{\citenamefont {Ren}\ \emph {et~al.}(2023)\citenamefont {Ren},
  \citenamefont {Fu}, \citenamefont {Wu},\ and\ \citenamefont
  {Chen}}]{ren2023towards}%
  \BibitemOpen
  \bibfield  {author} {\bibinfo {author} {\bibfnamefont {W.}~\bibnamefont
  {Ren}}, \bibinfo {author} {\bibfnamefont {W.}~\bibnamefont {Fu}}, \bibinfo
  {author} {\bibfnamefont {X.}~\bibnamefont {Wu}},\ and\ \bibinfo {author}
  {\bibfnamefont {J.}~\bibnamefont {Chen}},\ }\bibfield  {title} {\bibinfo
  {title} {Towards the ground state of molecules via diffusion monte carlo on
  neural networks},\ }\href {https://doi.org/10.1038/s41467-023-37609-3}
  {\bibfield  {journal} {\bibinfo  {journal} {Nature Communications}\ }\textbf
  {\bibinfo {volume} {14}},\ \bibinfo {pages} {1860} (\bibinfo {year}
  {2023})}\BibitemShut {NoStop}%
\bibitem [{\citenamefont {Han}\ \emph {et~al.}(2019)\citenamefont {Han},
  \citenamefont {Zhang},\ and\ \citenamefont {Weinan}}]{han2019solving}%
  \BibitemOpen
  \bibfield  {author} {\bibinfo {author} {\bibfnamefont {J.}~\bibnamefont
  {Han}}, \bibinfo {author} {\bibfnamefont {L.}~\bibnamefont {Zhang}},\ and\
  \bibinfo {author} {\bibfnamefont {E.}~\bibnamefont {Weinan}},\ }\bibfield
  {title} {\bibinfo {title} {Solving many-electron schr{\"o}dinger equation
  using deep neural networks},\ }\href
  {https://doi.org/https://doi.org/10.1016/j.jcp.2019.108929} {\bibfield
  {journal} {\bibinfo  {journal} {Journal of Computational Physics}\ }\textbf
  {\bibinfo {volume} {399}},\ \bibinfo {pages} {108929} (\bibinfo {year}
  {2019})}\BibitemShut {NoStop}%
\bibitem [{\citenamefont {Pfau}\ \emph {et~al.}(2020)\citenamefont {Pfau},
  \citenamefont {Spencer}, \citenamefont {Matthews},\ and\ \citenamefont
  {Foulkes}}]{pfau2020ab}%
  \BibitemOpen
  \bibfield  {author} {\bibinfo {author} {\bibfnamefont {D.}~\bibnamefont
  {Pfau}}, \bibinfo {author} {\bibfnamefont {J.~S.}\ \bibnamefont {Spencer}},
  \bibinfo {author} {\bibfnamefont {A.~G. D.~G.}\ \bibnamefont {Matthews}},\
  and\ \bibinfo {author} {\bibfnamefont {W.~M.~C.}\ \bibnamefont {Foulkes}},\
  }\bibfield  {title} {\bibinfo {title} {Ab initio solution of the
  many-electron schr\"odinger equation with deep neural networks},\ }\href
  {https://doi.org/10.1103/PhysRevResearch.2.033429} {\bibfield  {journal}
  {\bibinfo  {journal} {Phys. Rev. Res.}\ }\textbf {\bibinfo {volume} {2}},\
  \bibinfo {pages} {033429} (\bibinfo {year} {2020})}\BibitemShut {NoStop}%
\bibitem [{\citenamefont {Westerhout}\ \emph {et~al.}(2020)\citenamefont
  {Westerhout}, \citenamefont {Astrakhantsev}, \citenamefont {Tikhonov},
  \citenamefont {Katsnelson},\ and\ \citenamefont
  {Bagrov}}]{westerhout2020generalization}%
  \BibitemOpen
  \bibfield  {author} {\bibinfo {author} {\bibfnamefont {T.}~\bibnamefont
  {Westerhout}}, \bibinfo {author} {\bibfnamefont {N.}~\bibnamefont
  {Astrakhantsev}}, \bibinfo {author} {\bibfnamefont {K.~S.}\ \bibnamefont
  {Tikhonov}}, \bibinfo {author} {\bibfnamefont {M.~I.}\ \bibnamefont
  {Katsnelson}},\ and\ \bibinfo {author} {\bibfnamefont {A.~A.}\ \bibnamefont
  {Bagrov}},\ }\bibfield  {title} {\bibinfo {title} {Generalization properties
  of neural network approximations to frustrated magnet ground states},\ }\href
  {https://doi.org/10.1038/s41467-020-15402-w} {\bibfield  {journal} {\bibinfo
  {journal} {Nature communications}\ }\textbf {\bibinfo {volume} {11}},\
  \bibinfo {pages} {1593} (\bibinfo {year} {2020})}\BibitemShut {NoStop}%
\bibitem [{\citenamefont {Roth}\ \emph {et~al.}(2023)\citenamefont {Roth},
  \citenamefont {Szab\'o},\ and\ \citenamefont {MacDonald}}]{roth2023high}%
  \BibitemOpen
  \bibfield  {author} {\bibinfo {author} {\bibfnamefont {C.}~\bibnamefont
  {Roth}}, \bibinfo {author} {\bibfnamefont {A.}~\bibnamefont {Szab\'o}},\ and\
  \bibinfo {author} {\bibfnamefont {A.~H.}\ \bibnamefont {MacDonald}},\
  }\bibfield  {title} {\bibinfo {title} {High-accuracy variational monte carlo
  for frustrated magnets with deep neural networks},\ }\href
  {https://doi.org/10.1103/PhysRevB.108.054410} {\bibfield  {journal} {\bibinfo
   {journal} {Phys. Rev. B}\ }\textbf {\bibinfo {volume} {108}},\ \bibinfo
  {pages} {054410} (\bibinfo {year} {2023})}\BibitemShut {NoStop}%
\bibitem [{\citenamefont {Szab\'o}\ and\ \citenamefont
  {Castelnovo}(2020)}]{szabo2020neural}%
  \BibitemOpen
  \bibfield  {author} {\bibinfo {author} {\bibfnamefont {A.}~\bibnamefont
  {Szab\'o}}\ and\ \bibinfo {author} {\bibfnamefont {C.}~\bibnamefont
  {Castelnovo}},\ }\bibfield  {title} {\bibinfo {title} {Neural network wave
  functions and the sign problem},\ }\href
  {https://doi.org/10.1103/PhysRevResearch.2.033075} {\bibfield  {journal}
  {\bibinfo  {journal} {Phys. Rev. Res.}\ }\textbf {\bibinfo {volume} {2}},\
  \bibinfo {pages} {033075} (\bibinfo {year} {2020})}\BibitemShut {NoStop}%
\bibitem [{\citenamefont {Chen}\ \emph {et~al.}(2022)\citenamefont {Chen},
  \citenamefont {Choo}, \citenamefont {Astrakhantsev},\ and\ \citenamefont
  {Neupert}}]{chen2022neural}%
  \BibitemOpen
  \bibfield  {author} {\bibinfo {author} {\bibfnamefont {A.}~\bibnamefont
  {Chen}}, \bibinfo {author} {\bibfnamefont {K.}~\bibnamefont {Choo}}, \bibinfo
  {author} {\bibfnamefont {N.}~\bibnamefont {Astrakhantsev}},\ and\ \bibinfo
  {author} {\bibfnamefont {T.}~\bibnamefont {Neupert}},\ }\bibfield  {title}
  {\bibinfo {title} {Neural network evolution strategy for solving quantum sign
  structures},\ }\href {https://doi.org/10.1103/PhysRevResearch.4.L022026}
  {\bibfield  {journal} {\bibinfo  {journal} {Phys. Rev. Res.}\ }\textbf
  {\bibinfo {volume} {4}},\ \bibinfo {pages} {L022026} (\bibinfo {year}
  {2022})}\BibitemShut {NoStop}%
\bibitem [{\citenamefont {Schmitt}\ and\ \citenamefont
  {Heyl}(2020)}]{schmitt2020quantum}%
  \BibitemOpen
  \bibfield  {author} {\bibinfo {author} {\bibfnamefont {M.}~\bibnamefont
  {Schmitt}}\ and\ \bibinfo {author} {\bibfnamefont {M.}~\bibnamefont {Heyl}},\
  }\bibfield  {title} {\bibinfo {title} {Quantum many-body dynamics in two
  dimensions with artificial neural networks},\ }\href
  {https://doi.org/10.1103/PhysRevLett.125.100503} {\bibfield  {journal}
  {\bibinfo  {journal} {Phys. Rev. Lett.}\ }\textbf {\bibinfo {volume} {125}},\
  \bibinfo {pages} {100503} (\bibinfo {year} {2020})}\BibitemShut {NoStop}%
\bibitem [{\citenamefont {Guti{\'{e}}rrez}\ and\ \citenamefont
  {Mendl}(2022)}]{gutierrez2022real}%
  \BibitemOpen
  \bibfield  {author} {\bibinfo {author} {\bibfnamefont {I.~L.}\ \bibnamefont
  {Guti{\'{e}}rrez}}\ and\ \bibinfo {author} {\bibfnamefont {C.~B.}\
  \bibnamefont {Mendl}},\ }\bibfield  {title} {\bibinfo {title} {Real time
  evolution with neural-network quantum states},\ }\href
  {https://doi.org/10.22331/q-2022-01-20-627} {\bibfield  {journal} {\bibinfo
  {journal} {{Quantum}}\ }\textbf {\bibinfo {volume} {6}},\ \bibinfo {pages}
  {627} (\bibinfo {year} {2022})}\BibitemShut {NoStop}%
\bibitem [{\citenamefont {Sinibaldi}\ \emph
  {et~al.}(2023{\natexlab{a}})\citenamefont {Sinibaldi}, \citenamefont
  {Giuliani}, \citenamefont {Carleo},\ and\ \citenamefont
  {Vicentini}}]{SinibaldiVicentini2023}%
  \BibitemOpen
  \bibfield  {author} {\bibinfo {author} {\bibfnamefont {A.}~\bibnamefont
  {Sinibaldi}}, \bibinfo {author} {\bibfnamefont {C.}~\bibnamefont {Giuliani}},
  \bibinfo {author} {\bibfnamefont {G.}~\bibnamefont {Carleo}},\ and\ \bibinfo
  {author} {\bibfnamefont {F.}~\bibnamefont {Vicentini}},\ }\bibfield  {title}
  {\bibinfo {title} {Unbiasing time-dependent variational monte carlo by
  projected quantum evolution},\ }\href
  {https://quantum-journal.org/papers/q-2023-10-10-1131/} {\bibfield  {journal}
  {\bibinfo  {journal} {Quantum}\ }\textbf {\bibinfo {volume} {7}},\ \bibinfo
  {pages} {1131} (\bibinfo {year} {2023}{\natexlab{a}})}\BibitemShut {NoStop}%
\bibitem [{\citenamefont {Donatella}\ \emph {et~al.}(2023)\citenamefont
  {Donatella}, \citenamefont {Denis}, \citenamefont {Le~Boit\'e},\ and\
  \citenamefont {Ciuti}}]{donatella2023dynamics}%
  \BibitemOpen
  \bibfield  {author} {\bibinfo {author} {\bibfnamefont {K.}~\bibnamefont
  {Donatella}}, \bibinfo {author} {\bibfnamefont {Z.}~\bibnamefont {Denis}},
  \bibinfo {author} {\bibfnamefont {A.}~\bibnamefont {Le~Boit\'e}},\ and\
  \bibinfo {author} {\bibfnamefont {C.}~\bibnamefont {Ciuti}},\ }\bibfield
  {title} {\bibinfo {title} {Dynamics with autoregressive neural quantum
  states: Application to critical quench dynamics},\ }\href
  {https://doi.org/10.1103/PhysRevA.108.022210} {\bibfield  {journal} {\bibinfo
   {journal} {Phys. Rev. A}\ }\textbf {\bibinfo {volume} {108}},\ \bibinfo
  {pages} {022210} (\bibinfo {year} {2023})}\BibitemShut {NoStop}%
\bibitem [{\citenamefont {Burau}\ and\ \citenamefont
  {Heyl}(2021)}]{burau2021unitary}%
  \BibitemOpen
  \bibfield  {author} {\bibinfo {author} {\bibfnamefont {H.}~\bibnamefont
  {Burau}}\ and\ \bibinfo {author} {\bibfnamefont {M.}~\bibnamefont {Heyl}},\
  }\bibfield  {title} {\bibinfo {title} {Unitary long-time evolution with
  quantum renormalization groups and artificial neural networks},\ }\href
  {https://doi.org/10.1103/PhysRevLett.127.050601} {\bibfield  {journal}
  {\bibinfo  {journal} {Phys. Rev. Lett.}\ }\textbf {\bibinfo {volume} {127}},\
  \bibinfo {pages} {050601} (\bibinfo {year} {2021})}\BibitemShut {NoStop}%
\bibitem [{\citenamefont {Schmitt}\ \emph {et~al.}(2022)\citenamefont
  {Schmitt}, \citenamefont {Rams}, \citenamefont {Dziarmaga}, \citenamefont
  {Heyl},\ and\ \citenamefont {Zurek}}]{SchmittZurek2022}%
  \BibitemOpen
  \bibfield  {author} {\bibinfo {author} {\bibfnamefont {M.}~\bibnamefont
  {Schmitt}}, \bibinfo {author} {\bibfnamefont {M.~M.}\ \bibnamefont {Rams}},
  \bibinfo {author} {\bibfnamefont {J.}~\bibnamefont {Dziarmaga}}, \bibinfo
  {author} {\bibfnamefont {M.}~\bibnamefont {Heyl}},\ and\ \bibinfo {author}
  {\bibfnamefont {W.~H.}\ \bibnamefont {Zurek}},\ }\bibfield  {title} {\bibinfo
  {title} {Quantum phase transition dynamics in the two-dimensional
  transverse-field ising model},\ }\href
  {https://doi.org/10.1126/sciadv.abl6850} {\bibfield  {journal} {\bibinfo
  {journal} {Science Advances}\ }\textbf {\bibinfo {volume} {8}},\ \bibinfo
  {pages} {eabl6850} (\bibinfo {year} {2022})}\BibitemShut {NoStop}%
\bibitem [{\citenamefont {Vicentini}\ \emph {et~al.}(2019)\citenamefont
  {Vicentini}, \citenamefont {Biella}, \citenamefont {Regnault},\ and\
  \citenamefont {Ciuti}}]{vicentini2019variational}%
  \BibitemOpen
  \bibfield  {author} {\bibinfo {author} {\bibfnamefont {F.}~\bibnamefont
  {Vicentini}}, \bibinfo {author} {\bibfnamefont {A.}~\bibnamefont {Biella}},
  \bibinfo {author} {\bibfnamefont {N.}~\bibnamefont {Regnault}},\ and\
  \bibinfo {author} {\bibfnamefont {C.}~\bibnamefont {Ciuti}},\ }\bibfield
  {title} {\bibinfo {title} {Variational neural-network ansatz for steady
  states in open quantum systems},\ }\href
  {https://doi.org/10.1103/PhysRevLett.122.250503} {\bibfield  {journal}
  {\bibinfo  {journal} {Phys. Rev. Lett.}\ }\textbf {\bibinfo {volume} {122}},\
  \bibinfo {pages} {250503} (\bibinfo {year} {2019})}\BibitemShut {NoStop}%
\bibitem [{\citenamefont {Nagy}\ and\ \citenamefont
  {Savona}(2019)}]{nagy2019variational}%
  \BibitemOpen
  \bibfield  {author} {\bibinfo {author} {\bibfnamefont {A.}~\bibnamefont
  {Nagy}}\ and\ \bibinfo {author} {\bibfnamefont {V.}~\bibnamefont {Savona}},\
  }\bibfield  {title} {\bibinfo {title} {Variational quantum monte carlo method
  with a neural-network ansatz for open quantum systems},\ }\href
  {https://doi.org/10.1103/PhysRevLett.122.250501} {\bibfield  {journal}
  {\bibinfo  {journal} {Phys. Rev. Lett.}\ }\textbf {\bibinfo {volume} {122}},\
  \bibinfo {pages} {250501} (\bibinfo {year} {2019})}\BibitemShut {NoStop}%
\bibitem [{\citenamefont {Yoshioka}\ and\ \citenamefont
  {Hamazaki}(2019)}]{yoshioka2019constructing}%
  \BibitemOpen
  \bibfield  {author} {\bibinfo {author} {\bibfnamefont {N.}~\bibnamefont
  {Yoshioka}}\ and\ \bibinfo {author} {\bibfnamefont {R.}~\bibnamefont
  {Hamazaki}},\ }\bibfield  {title} {\bibinfo {title} {Constructing neural
  stationary states for open quantum many-body systems},\ }\href
  {https://doi.org/10.1103/PhysRevB.99.214306} {\bibfield  {journal} {\bibinfo
  {journal} {Phys. Rev. B}\ }\textbf {\bibinfo {volume} {99}},\ \bibinfo
  {pages} {214306} (\bibinfo {year} {2019})}\BibitemShut {NoStop}%
\bibitem [{\citenamefont {Hartmann}\ and\ \citenamefont
  {Carleo}(2019)}]{hartmann2019neural}%
  \BibitemOpen
  \bibfield  {author} {\bibinfo {author} {\bibfnamefont {M.~J.}\ \bibnamefont
  {Hartmann}}\ and\ \bibinfo {author} {\bibfnamefont {G.}~\bibnamefont
  {Carleo}},\ }\bibfield  {title} {\bibinfo {title} {Neural-network approach to
  dissipative quantum many-body dynamics},\ }\href
  {https://doi.org/10.1103/PhysRevLett.122.250502} {\bibfield  {journal}
  {\bibinfo  {journal} {Phys. Rev. Lett.}\ }\textbf {\bibinfo {volume} {122}},\
  \bibinfo {pages} {250502} (\bibinfo {year} {2019})}\BibitemShut {NoStop}%
\bibitem [{\citenamefont {Reh}\ \emph {et~al.}(2021)\citenamefont {Reh},
  \citenamefont {Schmitt},\ and\ \citenamefont {G\"arttner}}]{reh2021time}%
  \BibitemOpen
  \bibfield  {author} {\bibinfo {author} {\bibfnamefont {M.}~\bibnamefont
  {Reh}}, \bibinfo {author} {\bibfnamefont {M.}~\bibnamefont {Schmitt}},\ and\
  \bibinfo {author} {\bibfnamefont {M.}~\bibnamefont {G\"arttner}},\ }\bibfield
   {title} {\bibinfo {title} {Time-dependent variational principle for open
  quantum systems with artificial neural networks},\ }\href
  {https://doi.org/10.1103/PhysRevLett.127.230501} {\bibfield  {journal}
  {\bibinfo  {journal} {Phys. Rev. Lett.}\ }\textbf {\bibinfo {volume} {127}},\
  \bibinfo {pages} {230501} (\bibinfo {year} {2021})}\BibitemShut {NoStop}%
\bibitem [{\citenamefont {Carleo}\ \emph {et~al.}(2019)\citenamefont {Carleo},
  \citenamefont {Cirac}, \citenamefont {Cranmer}, \citenamefont {Daudet},
  \citenamefont {Schuld}, \citenamefont {Tishby}, \citenamefont
  {Vogt-Maranto},\ and\ \citenamefont {Zdeborov\'a}}]{carleo2019machine}%
  \BibitemOpen
  \bibfield  {author} {\bibinfo {author} {\bibfnamefont {G.}~\bibnamefont
  {Carleo}}, \bibinfo {author} {\bibfnamefont {I.}~\bibnamefont {Cirac}},
  \bibinfo {author} {\bibfnamefont {K.}~\bibnamefont {Cranmer}}, \bibinfo
  {author} {\bibfnamefont {L.}~\bibnamefont {Daudet}}, \bibinfo {author}
  {\bibfnamefont {M.}~\bibnamefont {Schuld}}, \bibinfo {author} {\bibfnamefont
  {N.}~\bibnamefont {Tishby}}, \bibinfo {author} {\bibfnamefont
  {L.}~\bibnamefont {Vogt-Maranto}},\ and\ \bibinfo {author} {\bibfnamefont
  {L.}~\bibnamefont {Zdeborov\'a}},\ }\bibfield  {title} {\bibinfo {title}
  {Machine learning and the physical sciences},\ }\href
  {https://doi.org/10.1103/RevModPhys.91.045002} {\bibfield  {journal}
  {\bibinfo  {journal} {Rev. Mod. Phys.}\ }\textbf {\bibinfo {volume} {91}},\
  \bibinfo {pages} {045002} (\bibinfo {year} {2019})}\BibitemShut {NoStop}%
\bibitem [{\citenamefont {Jia}\ \emph {et~al.}(2019)\citenamefont {Jia},
  \citenamefont {Yi}, \citenamefont {Zhai}, \citenamefont {Wu}, \citenamefont
  {Guo},\ and\ \citenamefont {Guo}}]{jia2019quantum}%
  \BibitemOpen
  \bibfield  {author} {\bibinfo {author} {\bibfnamefont {Z.-A.}\ \bibnamefont
  {Jia}}, \bibinfo {author} {\bibfnamefont {B.}~\bibnamefont {Yi}}, \bibinfo
  {author} {\bibfnamefont {R.}~\bibnamefont {Zhai}}, \bibinfo {author}
  {\bibfnamefont {Y.-C.}\ \bibnamefont {Wu}}, \bibinfo {author} {\bibfnamefont
  {G.-C.}\ \bibnamefont {Guo}},\ and\ \bibinfo {author} {\bibfnamefont {G.-P.}\
  \bibnamefont {Guo}},\ }\bibfield  {title} {\bibinfo {title} {Quantum neural
  network states: A brief review of methods and applications},\ }\href
  {https://doi.org/10.1002/qute.201800077} {\bibfield  {journal} {\bibinfo
  {journal} {Advanced Quantum Technologies}\ }\textbf {\bibinfo {volume} {2}},\
  \bibinfo {pages} {1800077} (\bibinfo {year} {2019})}\BibitemShut {NoStop}%
\bibitem [{\citenamefont {Carrasquilla}(2020)}]{carrasquilla2020machine}%
  \BibitemOpen
  \bibfield  {author} {\bibinfo {author} {\bibfnamefont {J.}~\bibnamefont
  {Carrasquilla}},\ }\bibfield  {title} {\bibinfo {title} {Machine learning for
  quantum matter},\ }\href {https://doi.org/10.1080/23746149.2020.1797528}
  {\bibfield  {journal} {\bibinfo  {journal} {Advances in Physics: X}\ }\textbf
  {\bibinfo {volume} {5}},\ \bibinfo {pages} {1797528} (\bibinfo {year}
  {2020})}\BibitemShut {NoStop}%
\bibitem [{\citenamefont {Guest}\ \emph {et~al.}(2018)\citenamefont {Guest},
  \citenamefont {Cranmer},\ and\ \citenamefont {Whiteson}}]{guest2018deep}%
  \BibitemOpen
  \bibfield  {author} {\bibinfo {author} {\bibfnamefont {D.}~\bibnamefont
  {Guest}}, \bibinfo {author} {\bibfnamefont {K.}~\bibnamefont {Cranmer}},\
  and\ \bibinfo {author} {\bibfnamefont {D.}~\bibnamefont {Whiteson}},\
  }\bibfield  {title} {\bibinfo {title} {Deep learning and its application to
  lhc physics},\ }\href {https://doi.org/10.1146/annurev-nucl-101917-021019}
  {\bibfield  {journal} {\bibinfo  {journal} {Annual Review of Nuclear and
  Particle Science}\ }\textbf {\bibinfo {volume} {68}},\ \bibinfo {pages} {161}
  (\bibinfo {year} {2018})}\BibitemShut {NoStop}%
\bibitem [{\citenamefont {Larkoski}\ \emph {et~al.}(2020)\citenamefont
  {Larkoski}, \citenamefont {Moult},\ and\ \citenamefont
  {Nachman}}]{larkoski2020jet}%
  \BibitemOpen
  \bibfield  {author} {\bibinfo {author} {\bibfnamefont {A.~J.}\ \bibnamefont
  {Larkoski}}, \bibinfo {author} {\bibfnamefont {I.}~\bibnamefont {Moult}},\
  and\ \bibinfo {author} {\bibfnamefont {B.}~\bibnamefont {Nachman}},\
  }\bibfield  {title} {\bibinfo {title} {Jet substructure at the large hadron
  collider: a review of recent advances in theory and machine learning},\
  }\href {https://doi.org/https://doi.org/10.1016/j.physrep.2019.11.001}
  {\bibfield  {journal} {\bibinfo  {journal} {Physics Reports}\ }\textbf
  {\bibinfo {volume} {841}},\ \bibinfo {pages} {1} (\bibinfo {year}
  {2020})}\BibitemShut {NoStop}%
\bibitem [{\citenamefont {Smolensky}\ \emph {et~al.}(1986)\citenamefont
  {Smolensky} \emph {et~al.}}]{smolensky1986information}%
  \BibitemOpen
  \bibfield  {author} {\bibinfo {author} {\bibfnamefont {P.}~\bibnamefont
  {Smolensky}} \emph {et~al.},\ }\bibfield  {title} {\bibinfo {title}
  {Information processing in dynamical systems: Foundations of harmony theory}\
  }\href {https://doi.org/10.4324/9780429433054-1} {10.4324/9780429433054-1}
  (\bibinfo {year} {1986})\BibitemShut {NoStop}%
\bibitem [{\citenamefont {Nomura}\ \emph {et~al.}(2021)\citenamefont {Nomura},
  \citenamefont {Yoshioka},\ and\ \citenamefont {Nori}}]{nomura2021purifying}%
  \BibitemOpen
  \bibfield  {author} {\bibinfo {author} {\bibfnamefont {Y.}~\bibnamefont
  {Nomura}}, \bibinfo {author} {\bibfnamefont {N.}~\bibnamefont {Yoshioka}},\
  and\ \bibinfo {author} {\bibfnamefont {F.}~\bibnamefont {Nori}},\ }\bibfield
  {title} {\bibinfo {title} {Purifying deep boltzmann machines for thermal
  quantum states},\ }\href {https://doi.org/10.1103/PhysRevLett.127.060601}
  {\bibfield  {journal} {\bibinfo  {journal} {Phys. Rev. Lett.}\ }\textbf
  {\bibinfo {volume} {127}},\ \bibinfo {pages} {060601} (\bibinfo {year}
  {2021})}\BibitemShut {NoStop}%
\bibitem [{\citenamefont {Melko}\ \emph {et~al.}(2019)\citenamefont {Melko},
  \citenamefont {Carleo}, \citenamefont {Carrasquilla},\ and\ \citenamefont
  {Cirac}}]{melko2019restricted}%
  \BibitemOpen
  \bibfield  {author} {\bibinfo {author} {\bibfnamefont {R.~G.}\ \bibnamefont
  {Melko}}, \bibinfo {author} {\bibfnamefont {G.}~\bibnamefont {Carleo}},
  \bibinfo {author} {\bibfnamefont {J.}~\bibnamefont {Carrasquilla}},\ and\
  \bibinfo {author} {\bibfnamefont {J.~I.}\ \bibnamefont {Cirac}},\ }\bibfield
  {title} {\bibinfo {title} {Restricted boltzmann machines in quantum
  physics},\ }\href
  {https://doi.org/https://doi.org/10.1016/j.physrep.2019.11.001} {\bibfield
  {journal} {\bibinfo  {journal} {Nature Physics}\ }\textbf {\bibinfo {volume}
  {15}},\ \bibinfo {pages} {887} (\bibinfo {year} {2019})}\BibitemShut
  {NoStop}%
\bibitem [{\citenamefont {Golubeva}\ and\ \citenamefont
  {Melko}(2022)}]{golubeva2022pruning}%
  \BibitemOpen
  \bibfield  {author} {\bibinfo {author} {\bibfnamefont {A.}~\bibnamefont
  {Golubeva}}\ and\ \bibinfo {author} {\bibfnamefont {R.~G.}\ \bibnamefont
  {Melko}},\ }\bibfield  {title} {\bibinfo {title} {Pruning a restricted
  boltzmann machine for quantum state reconstruction},\ }\href
  {https://doi.org/10.1103/PhysRevB.105.125124} {\bibfield  {journal} {\bibinfo
   {journal} {Phys. Rev. B}\ }\textbf {\bibinfo {volume} {105}},\ \bibinfo
  {pages} {125124} (\bibinfo {year} {2022})}\BibitemShut {NoStop}%
\bibitem [{\citenamefont {Zen}\ \emph {et~al.}(2020)\citenamefont {Zen},
  \citenamefont {My}, \citenamefont {Tan}, \citenamefont {H\'ebert},
  \citenamefont {Gattobigio}, \citenamefont {Miniatura}, \citenamefont
  {Poletti},\ and\ \citenamefont {Bressan}}]{zen2020transfer}%
  \BibitemOpen
  \bibfield  {author} {\bibinfo {author} {\bibfnamefont {R.}~\bibnamefont
  {Zen}}, \bibinfo {author} {\bibfnamefont {L.}~\bibnamefont {My}}, \bibinfo
  {author} {\bibfnamefont {R.}~\bibnamefont {Tan}}, \bibinfo {author}
  {\bibfnamefont {F.}~\bibnamefont {H\'ebert}}, \bibinfo {author}
  {\bibfnamefont {M.}~\bibnamefont {Gattobigio}}, \bibinfo {author}
  {\bibfnamefont {C.}~\bibnamefont {Miniatura}}, \bibinfo {author}
  {\bibfnamefont {D.}~\bibnamefont {Poletti}},\ and\ \bibinfo {author}
  {\bibfnamefont {S.}~\bibnamefont {Bressan}},\ }\bibfield  {title} {\bibinfo
  {title} {Transfer learning for scalability of neural-network quantum
  states},\ }\href {https://doi.org/10.1103/PhysRevE.101.053301} {\bibfield
  {journal} {\bibinfo  {journal} {Phys. Rev. E}\ }\textbf {\bibinfo {volume}
  {101}},\ \bibinfo {pages} {053301} (\bibinfo {year} {2020})}\BibitemShut
  {NoStop}%
\bibitem [{\citenamefont {Park}\ and\ \citenamefont
  {Kastoryano}(2022)}]{park2022expressive}%
  \BibitemOpen
  \bibfield  {author} {\bibinfo {author} {\bibfnamefont {C.-Y.}\ \bibnamefont
  {Park}}\ and\ \bibinfo {author} {\bibfnamefont {M.~J.}\ \bibnamefont
  {Kastoryano}},\ }\bibfield  {title} {\bibinfo {title} {Expressive power of
  complex-valued restricted boltzmann machines for solving nonstoquastic
  hamiltonians},\ }\href {https://doi.org/10.1103/PhysRevB.106.134437}
  {\bibfield  {journal} {\bibinfo  {journal} {Phys. Rev. B}\ }\textbf {\bibinfo
  {volume} {106}},\ \bibinfo {pages} {134437} (\bibinfo {year}
  {2022})}\BibitemShut {NoStop}%
\bibitem [{\citenamefont {Lu}\ \emph {et~al.}(2019)\citenamefont {Lu},
  \citenamefont {Gao},\ and\ \citenamefont {Duan}}]{lu2019efficient}%
  \BibitemOpen
  \bibfield  {author} {\bibinfo {author} {\bibfnamefont {S.}~\bibnamefont
  {Lu}}, \bibinfo {author} {\bibfnamefont {X.}~\bibnamefont {Gao}},\ and\
  \bibinfo {author} {\bibfnamefont {L.-M.}\ \bibnamefont {Duan}},\ }\bibfield
  {title} {\bibinfo {title} {Efficient representation of topologically ordered
  states with restricted boltzmann machines},\ }\href
  {https://doi.org/10.1103/PhysRevB.99.155136} {\bibfield  {journal} {\bibinfo
  {journal} {Phys. Rev. B}\ }\textbf {\bibinfo {volume} {99}},\ \bibinfo
  {pages} {155136} (\bibinfo {year} {2019})}\BibitemShut {NoStop}%
\bibitem [{\citenamefont {Park}\ and\ \citenamefont
  {Kastoryano}(2020)}]{park2020geometry}%
  \BibitemOpen
  \bibfield  {author} {\bibinfo {author} {\bibfnamefont {C.-Y.}\ \bibnamefont
  {Park}}\ and\ \bibinfo {author} {\bibfnamefont {M.~J.}\ \bibnamefont
  {Kastoryano}},\ }\bibfield  {title} {\bibinfo {title} {Geometry of learning
  neural quantum states},\ }\href
  {https://doi.org/10.1103/PhysRevResearch.2.023232} {\bibfield  {journal}
  {\bibinfo  {journal} {Phys. Rev. Res.}\ }\textbf {\bibinfo {volume} {2}},\
  \bibinfo {pages} {023232} (\bibinfo {year} {2020})}\BibitemShut {NoStop}%
\bibitem [{\citenamefont {Nomura}\ \emph {et~al.}(2017)\citenamefont {Nomura},
  \citenamefont {Darmawan}, \citenamefont {Yamaji},\ and\ \citenamefont
  {Imada}}]{nomura2017restricted}%
  \BibitemOpen
  \bibfield  {author} {\bibinfo {author} {\bibfnamefont {Y.}~\bibnamefont
  {Nomura}}, \bibinfo {author} {\bibfnamefont {A.~S.}\ \bibnamefont
  {Darmawan}}, \bibinfo {author} {\bibfnamefont {Y.}~\bibnamefont {Yamaji}},\
  and\ \bibinfo {author} {\bibfnamefont {M.}~\bibnamefont {Imada}},\ }\bibfield
   {title} {\bibinfo {title} {Restricted boltzmann machine learning for solving
  strongly correlated quantum systems},\ }\href
  {https://doi.org/10.1103/PhysRevB.96.205152} {\bibfield  {journal} {\bibinfo
  {journal} {Phys. Rev. B}\ }\textbf {\bibinfo {volume} {96}},\ \bibinfo
  {pages} {205152} (\bibinfo {year} {2017})}\BibitemShut {NoStop}%
\bibitem [{\citenamefont {Nomura}(2021)}]{nomura2021helping}%
  \BibitemOpen
  \bibfield  {author} {\bibinfo {author} {\bibfnamefont {Y.}~\bibnamefont
  {Nomura}},\ }\bibfield  {title} {\bibinfo {title} {Helping restricted
  boltzmann machines with quantum-state representation by restoring symmetry},\
  }\href {https://doi.org/10.1088/1361-648X/abe268} {\bibfield  {journal}
  {\bibinfo  {journal} {Journal of Physics: Condensed Matter}\ }\textbf
  {\bibinfo {volume} {33}},\ \bibinfo {pages} {174003} (\bibinfo {year}
  {2021})}\BibitemShut {NoStop}%
\bibitem [{\citenamefont {Deng}\ \emph {et~al.}(2017)\citenamefont {Deng},
  \citenamefont {Li},\ and\ \citenamefont {Das~Sarma}}]{deng2017quantum}%
  \BibitemOpen
  \bibfield  {author} {\bibinfo {author} {\bibfnamefont {D.-L.}\ \bibnamefont
  {Deng}}, \bibinfo {author} {\bibfnamefont {X.}~\bibnamefont {Li}},\ and\
  \bibinfo {author} {\bibfnamefont {S.}~\bibnamefont {Das~Sarma}},\ }\bibfield
  {title} {\bibinfo {title} {Quantum entanglement in neural network states},\
  }\href {https://doi.org/10.1103/PhysRevX.7.021021} {\bibfield  {journal}
  {\bibinfo  {journal} {Phys. Rev. X}\ }\textbf {\bibinfo {volume} {7}},\
  \bibinfo {pages} {021021} (\bibinfo {year} {2017})}\BibitemShut {NoStop}%
\bibitem [{\citenamefont {King}\ \emph {et~al.}(2024)\citenamefont {King},
  \citenamefont {Nocera}, \citenamefont {Rams}, \citenamefont {Dziarmaga},
  \citenamefont {Wiersema}, \citenamefont {Bernoudy}, \citenamefont {Raymond},
  \citenamefont {Kaushal}, \citenamefont {Heinsdorf}, \citenamefont {Harris},
  \citenamefont {Boothby}, \citenamefont {Altomare}, \citenamefont {Berkley},
  \citenamefont {Boschnak}, \citenamefont {Chern}, \citenamefont {Christiani},
  \citenamefont {Cibere}, \citenamefont {Connor}, \citenamefont {Dehn},
  \citenamefont {Deshpande}, \citenamefont {Ejtemaee}, \citenamefont {Farré},
  \citenamefont {Hamer}, \citenamefont {Hoskinson}, \citenamefont {Huang},
  \citenamefont {Johnson}, \citenamefont {Kortas}, \citenamefont {Ladizinsky},
  \citenamefont {Lai}, \citenamefont {Lanting}, \citenamefont {Li},
  \citenamefont {MacDonald}, \citenamefont {Marsden}, \citenamefont {McGeoch},
  \citenamefont {Molavi}, \citenamefont {Neufeld}, \citenamefont {Norouzpour},
  \citenamefont {Oh}, \citenamefont {Pasvolsky}, \citenamefont {Poitras},
  \citenamefont {Poulin-Lamarre}, \citenamefont {Prescott}, \citenamefont
  {Reis}, \citenamefont {Rich}, \citenamefont {Samani}, \citenamefont
  {Sheldan}, \citenamefont {Smirnov}, \citenamefont {Sterpka}, \citenamefont
  {Clavera}, \citenamefont {Tsai}, \citenamefont {Volkmann}, \citenamefont
  {Whiticar}, \citenamefont {Whittaker}, \citenamefont {Wilkinson},
  \citenamefont {Yao}, \citenamefont {Yi}, \citenamefont {Sandvik},
  \citenamefont {Alvarez}, \citenamefont {Melko}, \citenamefont {Carrasquilla},
  \citenamefont {Franz},\ and\ \citenamefont {Amin}}]{kingamin2024}%
  \BibitemOpen
  \bibfield  {author} {\bibinfo {author} {\bibfnamefont {A.~D.}\ \bibnamefont
  {King}}, \bibinfo {author} {\bibfnamefont {A.}~\bibnamefont {Nocera}},
  \bibinfo {author} {\bibfnamefont {M.~M.}\ \bibnamefont {Rams}}, \bibinfo
  {author} {\bibfnamefont {J.}~\bibnamefont {Dziarmaga}}, \bibinfo {author}
  {\bibfnamefont {R.}~\bibnamefont {Wiersema}}, \bibinfo {author}
  {\bibfnamefont {W.}~\bibnamefont {Bernoudy}}, \bibinfo {author}
  {\bibfnamefont {J.}~\bibnamefont {Raymond}}, \bibinfo {author} {\bibfnamefont
  {N.}~\bibnamefont {Kaushal}}, \bibinfo {author} {\bibfnamefont
  {N.}~\bibnamefont {Heinsdorf}}, \bibinfo {author} {\bibfnamefont
  {R.}~\bibnamefont {Harris}}, \bibinfo {author} {\bibfnamefont
  {K.}~\bibnamefont {Boothby}}, \bibinfo {author} {\bibfnamefont
  {F.}~\bibnamefont {Altomare}}, \bibinfo {author} {\bibfnamefont {A.~J.}\
  \bibnamefont {Berkley}}, \bibinfo {author} {\bibfnamefont {M.}~\bibnamefont
  {Boschnak}}, \bibinfo {author} {\bibfnamefont {K.}~\bibnamefont {Chern}},
  \bibinfo {author} {\bibfnamefont {H.}~\bibnamefont {Christiani}}, \bibinfo
  {author} {\bibfnamefont {S.}~\bibnamefont {Cibere}}, \bibinfo {author}
  {\bibfnamefont {J.}~\bibnamefont {Connor}}, \bibinfo {author} {\bibfnamefont
  {M.~H.}\ \bibnamefont {Dehn}}, \bibinfo {author} {\bibfnamefont
  {R.}~\bibnamefont {Deshpande}}, \bibinfo {author} {\bibfnamefont
  {S.}~\bibnamefont {Ejtemaee}}, \bibinfo {author} {\bibfnamefont
  {P.}~\bibnamefont {Farré}}, \bibinfo {author} {\bibfnamefont
  {K.}~\bibnamefont {Hamer}}, \bibinfo {author} {\bibfnamefont
  {E.}~\bibnamefont {Hoskinson}}, \bibinfo {author} {\bibfnamefont
  {S.}~\bibnamefont {Huang}}, \bibinfo {author} {\bibfnamefont {M.~W.}\
  \bibnamefont {Johnson}}, \bibinfo {author} {\bibfnamefont {S.}~\bibnamefont
  {Kortas}}, \bibinfo {author} {\bibfnamefont {E.}~\bibnamefont {Ladizinsky}},
  \bibinfo {author} {\bibfnamefont {T.}~\bibnamefont {Lai}}, \bibinfo {author}
  {\bibfnamefont {T.}~\bibnamefont {Lanting}}, \bibinfo {author} {\bibfnamefont
  {R.}~\bibnamefont {Li}}, \bibinfo {author} {\bibfnamefont {A.~J.~R.}\
  \bibnamefont {MacDonald}}, \bibinfo {author} {\bibfnamefont {G.}~\bibnamefont
  {Marsden}}, \bibinfo {author} {\bibfnamefont {C.~C.}\ \bibnamefont
  {McGeoch}}, \bibinfo {author} {\bibfnamefont {R.}~\bibnamefont {Molavi}},
  \bibinfo {author} {\bibfnamefont {R.}~\bibnamefont {Neufeld}}, \bibinfo
  {author} {\bibfnamefont {M.}~\bibnamefont {Norouzpour}}, \bibinfo {author}
  {\bibfnamefont {T.}~\bibnamefont {Oh}}, \bibinfo {author} {\bibfnamefont
  {J.}~\bibnamefont {Pasvolsky}}, \bibinfo {author} {\bibfnamefont
  {P.}~\bibnamefont {Poitras}}, \bibinfo {author} {\bibfnamefont
  {G.}~\bibnamefont {Poulin-Lamarre}}, \bibinfo {author} {\bibfnamefont
  {T.}~\bibnamefont {Prescott}}, \bibinfo {author} {\bibfnamefont
  {M.}~\bibnamefont {Reis}}, \bibinfo {author} {\bibfnamefont {C.}~\bibnamefont
  {Rich}}, \bibinfo {author} {\bibfnamefont {M.}~\bibnamefont {Samani}},
  \bibinfo {author} {\bibfnamefont {B.}~\bibnamefont {Sheldan}}, \bibinfo
  {author} {\bibfnamefont {A.}~\bibnamefont {Smirnov}}, \bibinfo {author}
  {\bibfnamefont {E.}~\bibnamefont {Sterpka}}, \bibinfo {author} {\bibfnamefont
  {B.~T.}\ \bibnamefont {Clavera}}, \bibinfo {author} {\bibfnamefont
  {N.}~\bibnamefont {Tsai}}, \bibinfo {author} {\bibfnamefont {M.}~\bibnamefont
  {Volkmann}}, \bibinfo {author} {\bibfnamefont {A.}~\bibnamefont {Whiticar}},
  \bibinfo {author} {\bibfnamefont {J.~D.}\ \bibnamefont {Whittaker}}, \bibinfo
  {author} {\bibfnamefont {W.}~\bibnamefont {Wilkinson}}, \bibinfo {author}
  {\bibfnamefont {J.}~\bibnamefont {Yao}}, \bibinfo {author} {\bibfnamefont
  {T.~J.}\ \bibnamefont {Yi}}, \bibinfo {author} {\bibfnamefont {A.~W.}\
  \bibnamefont {Sandvik}}, \bibinfo {author} {\bibfnamefont {G.}~\bibnamefont
  {Alvarez}}, \bibinfo {author} {\bibfnamefont {R.~G.}\ \bibnamefont {Melko}},
  \bibinfo {author} {\bibfnamefont {J.}~\bibnamefont {Carrasquilla}}, \bibinfo
  {author} {\bibfnamefont {M.}~\bibnamefont {Franz}},\ and\ \bibinfo {author}
  {\bibfnamefont {M.~H.}\ \bibnamefont {Amin}},\ }\href@noop {} {\bibinfo
  {title} {Computational supremacy in quantum simulation}} (\bibinfo {year}
  {2024}),\ \Eprint {https://arxiv.org/abs/2403.00910} {arXiv:2403.00910
  [quant-ph]} \BibitemShut {NoStop}%
\bibitem [{\citenamefont {Choo}\ \emph {et~al.}(2018)\citenamefont {Choo},
  \citenamefont {Carleo}, \citenamefont {Regnault},\ and\ \citenamefont
  {Neupert}}]{choo2018symmetries}%
  \BibitemOpen
  \bibfield  {author} {\bibinfo {author} {\bibfnamefont {K.}~\bibnamefont
  {Choo}}, \bibinfo {author} {\bibfnamefont {G.}~\bibnamefont {Carleo}},
  \bibinfo {author} {\bibfnamefont {N.}~\bibnamefont {Regnault}},\ and\
  \bibinfo {author} {\bibfnamefont {T.}~\bibnamefont {Neupert}},\ }\bibfield
  {title} {\bibinfo {title} {Symmetries and many-body excitations with
  neural-network quantum states},\ }\href
  {https://doi.org/10.1103/PhysRevLett.121.167204} {\bibfield  {journal}
  {\bibinfo  {journal} {Phys. Rev. Lett.}\ }\textbf {\bibinfo {volume} {121}},\
  \bibinfo {pages} {167204} (\bibinfo {year} {2018})}\BibitemShut {NoStop}%
\bibitem [{\citenamefont {Cai}\ and\ \citenamefont
  {Liu}(2018)}]{cai2018approximating}%
  \BibitemOpen
  \bibfield  {author} {\bibinfo {author} {\bibfnamefont {Z.}~\bibnamefont
  {Cai}}\ and\ \bibinfo {author} {\bibfnamefont {J.}~\bibnamefont {Liu}},\
  }\bibfield  {title} {\bibinfo {title} {Approximating quantum many-body wave
  functions using artificial neural networks},\ }\href
  {https://doi.org/10.1103/PhysRevB.97.035116} {\bibfield  {journal} {\bibinfo
  {journal} {Phys. Rev. B}\ }\textbf {\bibinfo {volume} {97}},\ \bibinfo
  {pages} {035116} (\bibinfo {year} {2018})}\BibitemShut {NoStop}%
\bibitem [{\citenamefont {Saito}(2017)}]{saito2017solving}%
  \BibitemOpen
  \bibfield  {author} {\bibinfo {author} {\bibfnamefont {H.}~\bibnamefont
  {Saito}},\ }\bibfield  {title} {\bibinfo {title} {Solving the bose--hubbard
  model with machine learning},\ }\href
  {https://doi.org/https://doi.org/10.7566/JPSJ.86.093001} {\bibfield
  {journal} {\bibinfo  {journal} {Journal of the Physical Society of Japan}\
  }\textbf {\bibinfo {volume} {86}},\ \bibinfo {pages} {093001} (\bibinfo
  {year} {2017})}\BibitemShut {NoStop}%
\bibitem [{\citenamefont {Beer}\ \emph {et~al.}(2020)\citenamefont {Beer},
  \citenamefont {Bondarenko}, \citenamefont {Farrelly}, \citenamefont
  {Osborne}, \citenamefont {Salzmann}, \citenamefont {Scheiermann},\ and\
  \citenamefont {Wolf}}]{beer2020training}%
  \BibitemOpen
  \bibfield  {author} {\bibinfo {author} {\bibfnamefont {K.}~\bibnamefont
  {Beer}}, \bibinfo {author} {\bibfnamefont {D.}~\bibnamefont {Bondarenko}},
  \bibinfo {author} {\bibfnamefont {T.}~\bibnamefont {Farrelly}}, \bibinfo
  {author} {\bibfnamefont {T.~J.}\ \bibnamefont {Osborne}}, \bibinfo {author}
  {\bibfnamefont {R.}~\bibnamefont {Salzmann}}, \bibinfo {author}
  {\bibfnamefont {D.}~\bibnamefont {Scheiermann}},\ and\ \bibinfo {author}
  {\bibfnamefont {R.}~\bibnamefont {Wolf}},\ }\bibfield  {title} {\bibinfo
  {title} {Training deep quantum neural networks},\ }\href
  {https://doi.org/10.1038/s41467-020-14454-2} {\bibfield  {journal} {\bibinfo
  {journal} {Nature communications}\ }\textbf {\bibinfo {volume} {11}},\
  \bibinfo {pages} {808} (\bibinfo {year} {2020})}\BibitemShut {NoStop}%
\bibitem [{\citenamefont {Inui}\ \emph {et~al.}(2021)\citenamefont {Inui},
  \citenamefont {Kato},\ and\ \citenamefont {Motome}}]{inui2021determinant}%
  \BibitemOpen
  \bibfield  {author} {\bibinfo {author} {\bibfnamefont {K.}~\bibnamefont
  {Inui}}, \bibinfo {author} {\bibfnamefont {Y.}~\bibnamefont {Kato}},\ and\
  \bibinfo {author} {\bibfnamefont {Y.}~\bibnamefont {Motome}},\ }\bibfield
  {title} {\bibinfo {title} {Determinant-free fermionic wave function using
  feed-forward neural networks},\ }\href
  {https://doi.org/10.1103/PhysRevResearch.3.043126} {\bibfield  {journal}
  {\bibinfo  {journal} {Phys. Rev. Res.}\ }\textbf {\bibinfo {volume} {3}},\
  \bibinfo {pages} {043126} (\bibinfo {year} {2021})}\BibitemShut {NoStop}%
\bibitem [{\citenamefont {Yang}\ \emph {et~al.}(2020)\citenamefont {Yang},
  \citenamefont {Leng}, \citenamefont {Yu}, \citenamefont {Patel},
  \citenamefont {Hu},\ and\ \citenamefont {Pu}}]{yang2020deep}%
  \BibitemOpen
  \bibfield  {author} {\bibinfo {author} {\bibfnamefont {L.}~\bibnamefont
  {Yang}}, \bibinfo {author} {\bibfnamefont {Z.}~\bibnamefont {Leng}}, \bibinfo
  {author} {\bibfnamefont {G.}~\bibnamefont {Yu}}, \bibinfo {author}
  {\bibfnamefont {A.}~\bibnamefont {Patel}}, \bibinfo {author} {\bibfnamefont
  {W.-J.}\ \bibnamefont {Hu}},\ and\ \bibinfo {author} {\bibfnamefont
  {H.}~\bibnamefont {Pu}},\ }\bibfield  {title} {\bibinfo {title} {Deep
  learning-enhanced variational monte carlo method for quantum many-body
  physics},\ }\href {https://doi.org/10.1103/PhysRevResearch.2.012039}
  {\bibfield  {journal} {\bibinfo  {journal} {Phys. Rev. Res.}\ }\textbf
  {\bibinfo {volume} {2}},\ \bibinfo {pages} {012039} (\bibinfo {year}
  {2020})}\BibitemShut {NoStop}%
\bibitem [{\citenamefont {Irikura}\ and\ \citenamefont
  {Saito}(2020)}]{irikura2020neural}%
  \BibitemOpen
  \bibfield  {author} {\bibinfo {author} {\bibfnamefont {N.}~\bibnamefont
  {Irikura}}\ and\ \bibinfo {author} {\bibfnamefont {H.}~\bibnamefont
  {Saito}},\ }\bibfield  {title} {\bibinfo {title} {Neural-network quantum
  states at finite temperature},\ }\href
  {https://doi.org/10.1103/PhysRevResearch.2.013284} {\bibfield  {journal}
  {\bibinfo  {journal} {Phys. Rev. Res.}\ }\textbf {\bibinfo {volume} {2}},\
  \bibinfo {pages} {013284} (\bibinfo {year} {2020})}\BibitemShut {NoStop}%
\bibitem [{\citenamefont {Vieijra}\ and\ \citenamefont
  {Nys}(2021)}]{vieijra2021many}%
  \BibitemOpen
  \bibfield  {author} {\bibinfo {author} {\bibfnamefont {T.}~\bibnamefont
  {Vieijra}}\ and\ \bibinfo {author} {\bibfnamefont {J.}~\bibnamefont {Nys}},\
  }\bibfield  {title} {\bibinfo {title} {Many-body quantum states with exact
  conservation of non-abelian and lattice symmetries through variational monte
  carlo},\ }\href {https://doi.org/10.1103/PhysRevB.104.045123} {\bibfield
  {journal} {\bibinfo  {journal} {Phys. Rev. B}\ }\textbf {\bibinfo {volume}
  {104}},\ \bibinfo {pages} {045123} (\bibinfo {year} {2021})}\BibitemShut
  {NoStop}%
\bibitem [{\citenamefont {Liang}\ \emph {et~al.}(2018)\citenamefont {Liang},
  \citenamefont {Liu}, \citenamefont {Lin}, \citenamefont {Guo}, \citenamefont
  {Zhang},\ and\ \citenamefont {He}}]{liang2018solving}%
  \BibitemOpen
  \bibfield  {author} {\bibinfo {author} {\bibfnamefont {X.}~\bibnamefont
  {Liang}}, \bibinfo {author} {\bibfnamefont {W.-Y.}\ \bibnamefont {Liu}},
  \bibinfo {author} {\bibfnamefont {P.-Z.}\ \bibnamefont {Lin}}, \bibinfo
  {author} {\bibfnamefont {G.-C.}\ \bibnamefont {Guo}}, \bibinfo {author}
  {\bibfnamefont {Y.-S.}\ \bibnamefont {Zhang}},\ and\ \bibinfo {author}
  {\bibfnamefont {L.}~\bibnamefont {He}},\ }\bibfield  {title} {\bibinfo
  {title} {Solving frustrated quantum many-particle models with convolutional
  neural networks},\ }\href {https://doi.org/10.1103/PhysRevB.98.104426}
  {\bibfield  {journal} {\bibinfo  {journal} {Phys. Rev. B}\ }\textbf {\bibinfo
  {volume} {98}},\ \bibinfo {pages} {104426} (\bibinfo {year}
  {2018})}\BibitemShut {NoStop}%
\bibitem [{\citenamefont {Liu}\ and\ \citenamefont
  {Wang}(2021)}]{liu2021random}%
  \BibitemOpen
  \bibfield  {author} {\bibinfo {author} {\bibfnamefont {C.-Y.}\ \bibnamefont
  {Liu}}\ and\ \bibinfo {author} {\bibfnamefont {D.-W.}\ \bibnamefont {Wang}},\
  }\bibfield  {title} {\bibinfo {title} {Random sampling neural network for
  quantum many-body problems},\ }\href
  {https://doi.org/10.1103/PhysRevB.103.205107} {\bibfield  {journal} {\bibinfo
   {journal} {Phys. Rev. B}\ }\textbf {\bibinfo {volume} {103}},\ \bibinfo
  {pages} {205107} (\bibinfo {year} {2021})}\BibitemShut {NoStop}%
\bibitem [{\citenamefont {Saito}\ and\ \citenamefont
  {Kato}(2018)}]{saito2018machine}%
  \BibitemOpen
  \bibfield  {author} {\bibinfo {author} {\bibfnamefont {H.}~\bibnamefont
  {Saito}}\ and\ \bibinfo {author} {\bibfnamefont {M.}~\bibnamefont {Kato}},\
  }\bibfield  {title} {\bibinfo {title} {Machine learning technique to find
  quantum many-body ground states of bosons on a lattice},\ }\href
  {https://doi.org/https://doi.org/10.7566/JPSJ.87.014001} {\bibfield
  {journal} {\bibinfo  {journal} {Journal of the Physical Society of Japan}\
  }\textbf {\bibinfo {volume} {87}},\ \bibinfo {pages} {014001} (\bibinfo
  {year} {2018})}\BibitemShut {NoStop}%
\bibitem [{\citenamefont {Roth}\ and\ \citenamefont
  {MacDonald}(2021)}]{roth2021group}%
  \BibitemOpen
  \bibfield  {author} {\bibinfo {author} {\bibfnamefont {C.}~\bibnamefont
  {Roth}}\ and\ \bibinfo {author} {\bibfnamefont {A.~H.}\ \bibnamefont
  {MacDonald}},\ }\href@noop {} {\bibinfo {title} {Group convolutional neural
  networks improve quantum state accuracy}} (\bibinfo {year} {2021}),\ \Eprint
  {https://arxiv.org/abs/2104.05085} {arXiv:2104.05085 [quant-ph]} \BibitemShut
  {NoStop}%
\bibitem [{\citenamefont {Fu}\ \emph {et~al.}()\citenamefont {Fu},
  \citenamefont {Zhang}, \citenamefont {Zhang}, \citenamefont {Ling},
  \citenamefont {Xu},\ and\ \citenamefont {Ji}}]{fu2024lattice}%
  \BibitemOpen
  \bibfield  {author} {\bibinfo {author} {\bibfnamefont {C.}~\bibnamefont
  {Fu}}, \bibinfo {author} {\bibfnamefont {X.}~\bibnamefont {Zhang}}, \bibinfo
  {author} {\bibfnamefont {H.}~\bibnamefont {Zhang}}, \bibinfo {author}
  {\bibfnamefont {H.}~\bibnamefont {Ling}}, \bibinfo {author} {\bibfnamefont
  {S.}~\bibnamefont {Xu}},\ and\ \bibinfo {author} {\bibfnamefont
  {S.}~\bibnamefont {Ji}},\ }\bibinfo {title} {Lattice convolutional networks
  for learning ground states of quantum many-body systems},\ in\ \href
  {https://doi.org/10.1137/1.9781611978032.57} {\emph {\bibinfo {booktitle}
  {Proceedings of the 2024 SIAM International Conference on Data Mining
  (SDM)}}},\ pp.\ \bibinfo {pages} {490--498}\BibitemShut {NoStop}%
\bibitem [{\citenamefont {Reh}\ \emph {et~al.}(2023)\citenamefont {Reh},
  \citenamefont {Schmitt},\ and\ \citenamefont
  {G\"arttner}}]{reh2023optimizing}%
  \BibitemOpen
  \bibfield  {author} {\bibinfo {author} {\bibfnamefont {M.}~\bibnamefont
  {Reh}}, \bibinfo {author} {\bibfnamefont {M.}~\bibnamefont {Schmitt}},\ and\
  \bibinfo {author} {\bibfnamefont {M.}~\bibnamefont {G\"arttner}},\ }\bibfield
   {title} {\bibinfo {title} {Optimizing design choices for neural quantum
  states},\ }\href {https://doi.org/10.1103/PhysRevB.107.195115} {\bibfield
  {journal} {\bibinfo  {journal} {Phys. Rev. B}\ }\textbf {\bibinfo {volume}
  {107}},\ \bibinfo {pages} {195115} (\bibinfo {year} {2023})}\BibitemShut
  {NoStop}%
\bibitem [{\citenamefont {Herrera~Rodríguez}\ and\ \citenamefont
  {Kananenka}(2021)}]{herrera2021convolutional}%
  \BibitemOpen
  \bibfield  {author} {\bibinfo {author} {\bibfnamefont {L.~E.}\ \bibnamefont
  {Herrera~Rodríguez}}\ and\ \bibinfo {author} {\bibfnamefont {A.~A.}\
  \bibnamefont {Kananenka}},\ }\bibfield  {title} {\bibinfo {title}
  {Convolutional neural networks for long time dissipative quantum dynamics},\
  }\href {https://doi.org/doi: 10.1021/acs.jpclett.1c00079} {\bibfield
  {journal} {\bibinfo  {journal} {The Journal of Physical Chemistry Letters}\
  }\textbf {\bibinfo {volume} {12}},\ \bibinfo {pages} {2476} (\bibinfo {year}
  {2021})}\BibitemShut {NoStop}%
\bibitem [{\citenamefont {Sharir}\ \emph {et~al.}(2020)\citenamefont {Sharir},
  \citenamefont {Levine}, \citenamefont {Wies}, \citenamefont {Carleo},\ and\
  \citenamefont {Shashua}}]{sharir2020deep}%
  \BibitemOpen
  \bibfield  {author} {\bibinfo {author} {\bibfnamefont {O.}~\bibnamefont
  {Sharir}}, \bibinfo {author} {\bibfnamefont {Y.}~\bibnamefont {Levine}},
  \bibinfo {author} {\bibfnamefont {N.}~\bibnamefont {Wies}}, \bibinfo {author}
  {\bibfnamefont {G.}~\bibnamefont {Carleo}},\ and\ \bibinfo {author}
  {\bibfnamefont {A.}~\bibnamefont {Shashua}},\ }\bibfield  {title} {\bibinfo
  {title} {Deep autoregressive models for the efficient variational simulation
  of many-body quantum systems},\ }\href
  {https://doi.org/10.1103/PhysRevLett.124.020503} {\bibfield  {journal}
  {\bibinfo  {journal} {Phys. Rev. Lett.}\ }\textbf {\bibinfo {volume} {124}},\
  \bibinfo {pages} {020503} (\bibinfo {year} {2020})}\BibitemShut {NoStop}%
\bibitem [{\citenamefont {Luo}\ \emph {et~al.}(2022)\citenamefont {Luo},
  \citenamefont {Chen}, \citenamefont {Carrasquilla},\ and\ \citenamefont
  {Clark}}]{luo2022autoregressive}%
  \BibitemOpen
  \bibfield  {author} {\bibinfo {author} {\bibfnamefont {D.}~\bibnamefont
  {Luo}}, \bibinfo {author} {\bibfnamefont {Z.}~\bibnamefont {Chen}}, \bibinfo
  {author} {\bibfnamefont {J.}~\bibnamefont {Carrasquilla}},\ and\ \bibinfo
  {author} {\bibfnamefont {B.~K.}\ \bibnamefont {Clark}},\ }\bibfield  {title}
  {\bibinfo {title} {Autoregressive neural network for simulating open quantum
  systems via a probabilistic formulation},\ }\href
  {https://doi.org/10.1103/PhysRevLett.128.090501} {\bibfield  {journal}
  {\bibinfo  {journal} {Phys. Rev. Lett.}\ }\textbf {\bibinfo {volume} {128}},\
  \bibinfo {pages} {090501} (\bibinfo {year} {2022})}\BibitemShut {NoStop}%
\bibitem [{\citenamefont {Zhang}\ and\ \citenamefont
  {Di~Ventra}(2023)}]{zhang2023transformer}%
  \BibitemOpen
  \bibfield  {author} {\bibinfo {author} {\bibfnamefont {Y.-H.}\ \bibnamefont
  {Zhang}}\ and\ \bibinfo {author} {\bibfnamefont {M.}~\bibnamefont
  {Di~Ventra}},\ }\bibfield  {title} {\bibinfo {title} {Transformer quantum
  state: A multipurpose model for quantum many-body problems},\ }\href
  {https://doi.org/10.1103/PhysRevB.107.075147} {\bibfield  {journal} {\bibinfo
   {journal} {Phys. Rev. B}\ }\textbf {\bibinfo {volume} {107}},\ \bibinfo
  {pages} {075147} (\bibinfo {year} {2023})}\BibitemShut {NoStop}%
\bibitem [{\citenamefont {Sorella}(1998)}]{Sorella1998}%
  \BibitemOpen
  \bibfield  {author} {\bibinfo {author} {\bibfnamefont {S.}~\bibnamefont
  {Sorella}},\ }\bibfield  {title} {\bibinfo {title} {Green function monte
  carlo with stochastic reconfiguration},\ }\href
  {https://doi.org/10.1103/PhysRevLett.80.4558} {\bibfield  {journal} {\bibinfo
   {journal} {Phys. Rev. Lett.}\ }\textbf {\bibinfo {volume} {80}},\ \bibinfo
  {pages} {4558} (\bibinfo {year} {1998})}\BibitemShut {NoStop}%
\bibitem [{\citenamefont {Vicentini}\ \emph {et~al.}(2022)\citenamefont
  {Vicentini}, \citenamefont {Hofmann}, \citenamefont {Szabó}, \citenamefont
  {Wu}, \citenamefont {Roth}, \citenamefont {Giuliani}, \citenamefont {Pescia},
  \citenamefont {Nys}, \citenamefont {Vargas-Calderón}, \citenamefont
  {Astrakhantsev},\ and\ \citenamefont {Carleo}}]{vicentini2022netket}%
  \BibitemOpen
  \bibfield  {author} {\bibinfo {author} {\bibfnamefont {F.}~\bibnamefont
  {Vicentini}}, \bibinfo {author} {\bibfnamefont {D.}~\bibnamefont {Hofmann}},
  \bibinfo {author} {\bibfnamefont {A.}~\bibnamefont {Szabó}}, \bibinfo
  {author} {\bibfnamefont {D.}~\bibnamefont {Wu}}, \bibinfo {author}
  {\bibfnamefont {C.}~\bibnamefont {Roth}}, \bibinfo {author} {\bibfnamefont
  {C.}~\bibnamefont {Giuliani}}, \bibinfo {author} {\bibfnamefont
  {G.}~\bibnamefont {Pescia}}, \bibinfo {author} {\bibfnamefont
  {J.}~\bibnamefont {Nys}}, \bibinfo {author} {\bibfnamefont {V.}~\bibnamefont
  {Vargas-Calderón}}, \bibinfo {author} {\bibfnamefont {N.}~\bibnamefont
  {Astrakhantsev}},\ and\ \bibinfo {author} {\bibfnamefont {G.}~\bibnamefont
  {Carleo}},\ }\bibfield  {title} {\bibinfo {title} {{NetKet 3: Machine
  Learning Toolbox for Many-Body Quantum Systems}},\ }\href
  {https://doi.org/10.21468/SciPostPhysCodeb.7} {\bibfield  {journal} {\bibinfo
   {journal} {SciPost Phys. Codebases}\ ,\ \bibinfo {pages} {7}} (\bibinfo
  {year} {2022})}\BibitemShut {NoStop}%
\bibitem [{\citenamefont {Martens}\ and\ \citenamefont
  {Grosse}(2015)}]{martens2015optimizing}%
  \BibitemOpen
  \bibfield  {author} {\bibinfo {author} {\bibfnamefont {J.}~\bibnamefont
  {Martens}}\ and\ \bibinfo {author} {\bibfnamefont {R.}~\bibnamefont
  {Grosse}},\ }\bibfield  {title} {\bibinfo {title} {Optimizing neural networks
  with kronecker-factored approximate curvature},\ }in\ \href
  {https://proceedings.mlr.press/v37/martens15.html} {\emph {\bibinfo
  {booktitle} {Proceedings of the 32nd International Conference on Machine
  Learning}}},\ \bibinfo {series} {Proceedings of Machine Learning Research},
  Vol.~\bibinfo {volume} {37},\ \bibinfo {editor} {edited by\ \bibinfo {editor}
  {\bibfnamefont {F.}~\bibnamefont {Bach}}\ and\ \bibinfo {editor}
  {\bibfnamefont {D.}~\bibnamefont {Blei}}}\ (\bibinfo  {publisher} {PMLR},\
  \bibinfo {address} {Lille, France},\ \bibinfo {year} {2015})\ pp.\ \bibinfo
  {pages} {2408--2417}\BibitemShut {NoStop}%
\bibitem [{\citenamefont {Zhang}\ \emph {et~al.}(2023)\citenamefont {Zhang},
  \citenamefont {Xu}, \citenamefont {Wu}, \citenamefont {Balachandran},\ and\
  \citenamefont {Poletti}}]{zhang2023ground}%
  \BibitemOpen
  \bibfield  {author} {\bibinfo {author} {\bibfnamefont {W.}~\bibnamefont
  {Zhang}}, \bibinfo {author} {\bibfnamefont {X.}~\bibnamefont {Xu}}, \bibinfo
  {author} {\bibfnamefont {Z.}~\bibnamefont {Wu}}, \bibinfo {author}
  {\bibfnamefont {V.}~\bibnamefont {Balachandran}},\ and\ \bibinfo {author}
  {\bibfnamefont {D.}~\bibnamefont {Poletti}},\ }\bibfield  {title} {\bibinfo
  {title} {Ground state search by local and sequential updates of neural
  network quantum states},\ }\href
  {https://doi.org/10.1103/PhysRevB.107.165149} {\bibfield  {journal} {\bibinfo
   {journal} {Phys. Rev. B}\ }\textbf {\bibinfo {volume} {107}},\ \bibinfo
  {pages} {165149} (\bibinfo {year} {2023})}\BibitemShut {NoStop}%
\bibitem [{\citenamefont {Chen}\ and\ \citenamefont
  {Heyl}(2023)}]{chen2023efficient}%
  \BibitemOpen
  \bibfield  {author} {\bibinfo {author} {\bibfnamefont {A.}~\bibnamefont
  {Chen}}\ and\ \bibinfo {author} {\bibfnamefont {M.}~\bibnamefont {Heyl}},\
  }\href@noop {} {\bibinfo {title} {Efficient optimization of deep neural
  quantum states toward machine precision}} (\bibinfo {year} {2023}),\ \Eprint
  {https://arxiv.org/abs/2302.01941} {arXiv:2302.01941 [cond-mat.dis-nn]}
  \BibitemShut {NoStop}%
\bibitem [{\citenamefont {Rende}\ \emph {et~al.}(2023)\citenamefont {Rende},
  \citenamefont {Viteritti}, \citenamefont {Bardone}, \citenamefont {Becca},\
  and\ \citenamefont {Goldt}}]{rende2023simple}%
  \BibitemOpen
  \bibfield  {author} {\bibinfo {author} {\bibfnamefont {R.}~\bibnamefont
  {Rende}}, \bibinfo {author} {\bibfnamefont {L.~L.}\ \bibnamefont
  {Viteritti}}, \bibinfo {author} {\bibfnamefont {L.}~\bibnamefont {Bardone}},
  \bibinfo {author} {\bibfnamefont {F.}~\bibnamefont {Becca}},\ and\ \bibinfo
  {author} {\bibfnamefont {S.}~\bibnamefont {Goldt}},\ }\href@noop {} {\bibinfo
  {title} {A simple linear algebra identity to optimize large-scale neural
  network quantum states}} (\bibinfo {year} {2023}),\ \Eprint
  {https://arxiv.org/abs/2310.05715} {arXiv:2310.05715 [cond-mat.str-el]}
  \BibitemShut {NoStop}%
\bibitem [{\citenamefont {Kollath}\ \emph {et~al.}(2007)\citenamefont
  {Kollath}, \citenamefont {L\"auchli},\ and\ \citenamefont
  {Altman}}]{kollath2007quench}%
  \BibitemOpen
  \bibfield  {author} {\bibinfo {author} {\bibfnamefont {C.}~\bibnamefont
  {Kollath}}, \bibinfo {author} {\bibfnamefont {A.~M.}\ \bibnamefont
  {L\"auchli}},\ and\ \bibinfo {author} {\bibfnamefont {E.}~\bibnamefont
  {Altman}},\ }\bibfield  {title} {\bibinfo {title} {Quench dynamics and
  nonequilibrium phase diagram of the bose-hubbard model},\ }\href
  {https://doi.org/10.1103/PhysRevLett.98.180601} {\bibfield  {journal}
  {\bibinfo  {journal} {Phys. Rev. Lett.}\ }\textbf {\bibinfo {volume} {98}},\
  \bibinfo {pages} {180601} (\bibinfo {year} {2007})}\BibitemShut {NoStop}%
\bibitem [{\citenamefont {Polkovnikov}\ \emph {et~al.}(2011)\citenamefont
  {Polkovnikov}, \citenamefont {Sengupta}, \citenamefont {Silva},\ and\
  \citenamefont {Vengalattore}}]{polkovnikov2011colloquium}%
  \BibitemOpen
  \bibfield  {author} {\bibinfo {author} {\bibfnamefont {A.}~\bibnamefont
  {Polkovnikov}}, \bibinfo {author} {\bibfnamefont {K.}~\bibnamefont
  {Sengupta}}, \bibinfo {author} {\bibfnamefont {A.}~\bibnamefont {Silva}},\
  and\ \bibinfo {author} {\bibfnamefont {M.}~\bibnamefont {Vengalattore}},\
  }\bibfield  {title} {\bibinfo {title} {Colloquium: Nonequilibrium dynamics of
  closed interacting quantum systems},\ }\href
  {https://doi.org/10.1103/RevModPhys.83.863} {\bibfield  {journal} {\bibinfo
  {journal} {Rev. Mod. Phys.}\ }\textbf {\bibinfo {volume} {83}},\ \bibinfo
  {pages} {863} (\bibinfo {year} {2011})}\BibitemShut {NoStop}%
\bibitem [{\citenamefont {Yukalov}(2011)}]{yukalov2011equilibration}%
  \BibitemOpen
  \bibfield  {author} {\bibinfo {author} {\bibfnamefont {V.~I.}\ \bibnamefont
  {Yukalov}},\ }\bibfield  {title} {\bibinfo {title} {Equilibration and
  thermalization in finite quantum systems},\ }\href
  {https://doi.org/10.1002/lapl.201110002} {\bibfield  {journal} {\bibinfo
  {journal} {Laser Physics Letters}\ }\textbf {\bibinfo {volume} {8}},\
  \bibinfo {pages} {485} (\bibinfo {year} {2011})}\BibitemShut {NoStop}%
\bibitem [{\citenamefont {Cassidy}\ \emph {et~al.}(2011)\citenamefont
  {Cassidy}, \citenamefont {Clark},\ and\ \citenamefont
  {Rigol}}]{cassidy2011generalized}%
  \BibitemOpen
  \bibfield  {author} {\bibinfo {author} {\bibfnamefont {A.~C.}\ \bibnamefont
  {Cassidy}}, \bibinfo {author} {\bibfnamefont {C.~W.}\ \bibnamefont {Clark}},\
  and\ \bibinfo {author} {\bibfnamefont {M.}~\bibnamefont {Rigol}},\ }\bibfield
   {title} {\bibinfo {title} {Generalized thermalization in an integrable
  lattice system},\ }\href {https://doi.org/10.1103/PhysRevLett.106.140405}
  {\bibfield  {journal} {\bibinfo  {journal} {Phys. Rev. Lett.}\ }\textbf
  {\bibinfo {volume} {106}},\ \bibinfo {pages} {140405} (\bibinfo {year}
  {2011})}\BibitemShut {NoStop}%
\bibitem [{\citenamefont {Canovi}\ \emph {et~al.}(2011)\citenamefont {Canovi},
  \citenamefont {Rossini}, \citenamefont {Fazio}, \citenamefont {Santoro},\
  and\ \citenamefont {Silva}}]{canovi2011quantum}%
  \BibitemOpen
  \bibfield  {author} {\bibinfo {author} {\bibfnamefont {E.}~\bibnamefont
  {Canovi}}, \bibinfo {author} {\bibfnamefont {D.}~\bibnamefont {Rossini}},
  \bibinfo {author} {\bibfnamefont {R.}~\bibnamefont {Fazio}}, \bibinfo
  {author} {\bibfnamefont {G.~E.}\ \bibnamefont {Santoro}},\ and\ \bibinfo
  {author} {\bibfnamefont {A.}~\bibnamefont {Silva}},\ }\bibfield  {title}
  {\bibinfo {title} {Quantum quenches, thermalization, and many-body
  localization},\ }\href {https://doi.org/10.1103/PhysRevB.83.094431}
  {\bibfield  {journal} {\bibinfo  {journal} {Phys. Rev. B}\ }\textbf {\bibinfo
  {volume} {83}},\ \bibinfo {pages} {094431} (\bibinfo {year}
  {2011})}\BibitemShut {NoStop}%
\bibitem [{\citenamefont {Rigol}(2009)}]{rigol2009quantum}%
  \BibitemOpen
  \bibfield  {author} {\bibinfo {author} {\bibfnamefont {M.}~\bibnamefont
  {Rigol}},\ }\bibfield  {title} {\bibinfo {title} {Quantum quenches and
  thermalization in one-dimensional fermionic systems},\ }\href
  {https://doi.org/10.1103/PhysRevA.80.053607} {\bibfield  {journal} {\bibinfo
  {journal} {Phys. Rev. A}\ }\textbf {\bibinfo {volume} {80}},\ \bibinfo
  {pages} {053607} (\bibinfo {year} {2009})}\BibitemShut {NoStop}%
\bibitem [{\citenamefont {Serbyn}\ \emph {et~al.}(2014)\citenamefont {Serbyn},
  \citenamefont {Papi\ifmmode~\acute{c}\else \'{c}\fi{}},\ and\ \citenamefont
  {Abanin}}]{serbyn2014quantum}%
  \BibitemOpen
  \bibfield  {author} {\bibinfo {author} {\bibfnamefont {M.}~\bibnamefont
  {Serbyn}}, \bibinfo {author} {\bibfnamefont {Z.}~\bibnamefont
  {Papi\ifmmode~\acute{c}\else \'{c}\fi{}}},\ and\ \bibinfo {author}
  {\bibfnamefont {D.~A.}\ \bibnamefont {Abanin}},\ }\bibfield  {title}
  {\bibinfo {title} {Quantum quenches in the many-body localized phase},\
  }\href {https://doi.org/10.1103/PhysRevB.90.174302} {\bibfield  {journal}
  {\bibinfo  {journal} {Phys. Rev. B}\ }\textbf {\bibinfo {volume} {90}},\
  \bibinfo {pages} {174302} (\bibinfo {year} {2014})}\BibitemShut {NoStop}%
\bibitem [{\citenamefont {Schreiber}\ \emph {et~al.}(2015)\citenamefont
  {Schreiber}, \citenamefont {Hodgman}, \citenamefont {Bordia}, \citenamefont
  {Lüschen}, \citenamefont {Fischer}, \citenamefont {Vosk}, \citenamefont
  {Altman}, \citenamefont {Schneider},\ and\ \citenamefont
  {Bloch}}]{schreiber2015observation}%
  \BibitemOpen
  \bibfield  {author} {\bibinfo {author} {\bibfnamefont {M.}~\bibnamefont
  {Schreiber}}, \bibinfo {author} {\bibfnamefont {S.~S.}\ \bibnamefont
  {Hodgman}}, \bibinfo {author} {\bibfnamefont {P.}~\bibnamefont {Bordia}},
  \bibinfo {author} {\bibfnamefont {H.~P.}\ \bibnamefont {Lüschen}}, \bibinfo
  {author} {\bibfnamefont {M.~H.}\ \bibnamefont {Fischer}}, \bibinfo {author}
  {\bibfnamefont {R.}~\bibnamefont {Vosk}}, \bibinfo {author} {\bibfnamefont
  {E.}~\bibnamefont {Altman}}, \bibinfo {author} {\bibfnamefont
  {U.}~\bibnamefont {Schneider}},\ and\ \bibinfo {author} {\bibfnamefont
  {I.}~\bibnamefont {Bloch}},\ }\bibfield  {title} {\bibinfo {title}
  {Observation of many-body localization of interacting fermions in a
  quasirandom optical lattice},\ }\href
  {https://doi.org/10.1126/science.aaa7432} {\bibfield  {journal} {\bibinfo
  {journal} {Science}\ }\textbf {\bibinfo {volume} {349}},\ \bibinfo {pages}
  {842} (\bibinfo {year} {2015})}\BibitemShut {NoStop}%
\bibitem [{\citenamefont {Kj\"all}\ \emph {et~al.}(2014)\citenamefont
  {Kj\"all}, \citenamefont {Bardarson},\ and\ \citenamefont
  {Pollmann}}]{kjall2014many}%
  \BibitemOpen
  \bibfield  {author} {\bibinfo {author} {\bibfnamefont {J.~A.}\ \bibnamefont
  {Kj\"all}}, \bibinfo {author} {\bibfnamefont {J.~H.}\ \bibnamefont
  {Bardarson}},\ and\ \bibinfo {author} {\bibfnamefont {F.}~\bibnamefont
  {Pollmann}},\ }\bibfield  {title} {\bibinfo {title} {Many-body localization
  in a disordered quantum ising chain},\ }\href
  {https://doi.org/10.1103/PhysRevLett.113.107204} {\bibfield  {journal}
  {\bibinfo  {journal} {Phys. Rev. Lett.}\ }\textbf {\bibinfo {volume} {113}},\
  \bibinfo {pages} {107204} (\bibinfo {year} {2014})}\BibitemShut {NoStop}%
\bibitem [{\citenamefont {Bardarson}\ \emph {et~al.}(2012)\citenamefont
  {Bardarson}, \citenamefont {Pollmann},\ and\ \citenamefont
  {Moore}}]{bardarson2012unbounded}%
  \BibitemOpen
  \bibfield  {author} {\bibinfo {author} {\bibfnamefont {J.~H.}\ \bibnamefont
  {Bardarson}}, \bibinfo {author} {\bibfnamefont {F.}~\bibnamefont
  {Pollmann}},\ and\ \bibinfo {author} {\bibfnamefont {J.~E.}\ \bibnamefont
  {Moore}},\ }\bibfield  {title} {\bibinfo {title} {Unbounded growth of
  entanglement in models of many-body localization},\ }\href
  {https://doi.org/10.1103/PhysRevLett.109.017202} {\bibfield  {journal}
  {\bibinfo  {journal} {Phys. Rev. Lett.}\ }\textbf {\bibinfo {volume} {109}},\
  \bibinfo {pages} {017202} (\bibinfo {year} {2012})}\BibitemShut {NoStop}%
\bibitem [{\citenamefont {Vosk}\ and\ \citenamefont
  {Altman}(2014)}]{vosk2014dynamical}%
  \BibitemOpen
  \bibfield  {author} {\bibinfo {author} {\bibfnamefont {R.}~\bibnamefont
  {Vosk}}\ and\ \bibinfo {author} {\bibfnamefont {E.}~\bibnamefont {Altman}},\
  }\bibfield  {title} {\bibinfo {title} {Dynamical quantum phase transitions in
  random spin chains},\ }\href {https://doi.org/10.1103/PhysRevLett.112.217204}
  {\bibfield  {journal} {\bibinfo  {journal} {Phys. Rev. Lett.}\ }\textbf
  {\bibinfo {volume} {112}},\ \bibinfo {pages} {217204} (\bibinfo {year}
  {2014})}\BibitemShut {NoStop}%
\bibitem [{\citenamefont {Serbyn}\ \emph {et~al.}(2021)\citenamefont {Serbyn},
  \citenamefont {Abanin},\ and\ \citenamefont {Papi{\'c}}}]{serbyn2021quantum}%
  \BibitemOpen
  \bibfield  {author} {\bibinfo {author} {\bibfnamefont {M.}~\bibnamefont
  {Serbyn}}, \bibinfo {author} {\bibfnamefont {D.~A.}\ \bibnamefont {Abanin}},\
  and\ \bibinfo {author} {\bibfnamefont {Z.}~\bibnamefont {Papi{\'c}}},\
  }\bibfield  {title} {\bibinfo {title} {Quantum many-body scars and weak
  breaking of ergodicity},\ }\href {https://doi.org/10.1038/s41567-021-01230-2}
  {\bibfield  {journal} {\bibinfo  {journal} {Nature Physics}\ }\textbf
  {\bibinfo {volume} {17}},\ \bibinfo {pages} {675} (\bibinfo {year}
  {2021})}\BibitemShut {NoStop}%
\bibitem [{\citenamefont {Su}\ \emph {et~al.}(2023)\citenamefont {Su},
  \citenamefont {Sun}, \citenamefont {Hudomal}, \citenamefont {Desaules},
  \citenamefont {Zhou}, \citenamefont {Yang}, \citenamefont {Halimeh},
  \citenamefont {Yuan}, \citenamefont {Papi\ifmmode~\acute{c}\else
  \'{c}\fi{}},\ and\ \citenamefont {Pan}}]{su2023observation}%
  \BibitemOpen
  \bibfield  {author} {\bibinfo {author} {\bibfnamefont {G.-X.}\ \bibnamefont
  {Su}}, \bibinfo {author} {\bibfnamefont {H.}~\bibnamefont {Sun}}, \bibinfo
  {author} {\bibfnamefont {A.}~\bibnamefont {Hudomal}}, \bibinfo {author}
  {\bibfnamefont {J.-Y.}\ \bibnamefont {Desaules}}, \bibinfo {author}
  {\bibfnamefont {Z.-Y.}\ \bibnamefont {Zhou}}, \bibinfo {author}
  {\bibfnamefont {B.}~\bibnamefont {Yang}}, \bibinfo {author} {\bibfnamefont
  {J.~C.}\ \bibnamefont {Halimeh}}, \bibinfo {author} {\bibfnamefont {Z.-S.}\
  \bibnamefont {Yuan}}, \bibinfo {author} {\bibfnamefont {Z.}~\bibnamefont
  {Papi\ifmmode~\acute{c}\else \'{c}\fi{}}},\ and\ \bibinfo {author}
  {\bibfnamefont {J.-W.}\ \bibnamefont {Pan}},\ }\bibfield  {title} {\bibinfo
  {title} {Observation of many-body scarring in a bose-hubbard quantum
  simulator},\ }\href {https://doi.org/10.1103/PhysRevResearch.5.023010}
  {\bibfield  {journal} {\bibinfo  {journal} {Phys. Rev. Res.}\ }\textbf
  {\bibinfo {volume} {5}},\ \bibinfo {pages} {023010} (\bibinfo {year}
  {2023})}\BibitemShut {NoStop}%
\bibitem [{\citenamefont {Schecter}\ and\ \citenamefont
  {Iadecola}(2019)}]{schecter2019weak}%
  \BibitemOpen
  \bibfield  {author} {\bibinfo {author} {\bibfnamefont {M.}~\bibnamefont
  {Schecter}}\ and\ \bibinfo {author} {\bibfnamefont {T.}~\bibnamefont
  {Iadecola}},\ }\bibfield  {title} {\bibinfo {title} {Weak ergodicity breaking
  and quantum many-body scars in spin-1 $xy$ magnets},\ }\href
  {https://doi.org/10.1103/PhysRevLett.123.147201} {\bibfield  {journal}
  {\bibinfo  {journal} {Phys. Rev. Lett.}\ }\textbf {\bibinfo {volume} {123}},\
  \bibinfo {pages} {147201} (\bibinfo {year} {2019})}\BibitemShut {NoStop}%
\bibitem [{\citenamefont {Ho}\ \emph {et~al.}(2019)\citenamefont {Ho},
  \citenamefont {Choi}, \citenamefont {Pichler},\ and\ \citenamefont
  {Lukin}}]{ho2019periodic}%
  \BibitemOpen
  \bibfield  {author} {\bibinfo {author} {\bibfnamefont {W.~W.}\ \bibnamefont
  {Ho}}, \bibinfo {author} {\bibfnamefont {S.}~\bibnamefont {Choi}}, \bibinfo
  {author} {\bibfnamefont {H.}~\bibnamefont {Pichler}},\ and\ \bibinfo {author}
  {\bibfnamefont {M.~D.}\ \bibnamefont {Lukin}},\ }\bibfield  {title} {\bibinfo
  {title} {Periodic orbits, entanglement, and quantum many-body scars in
  constrained models: Matrix product state approach},\ }\href
  {https://doi.org/10.1103/PhysRevLett.122.040603} {\bibfield  {journal}
  {\bibinfo  {journal} {Phys. Rev. Lett.}\ }\textbf {\bibinfo {volume} {122}},\
  \bibinfo {pages} {040603} (\bibinfo {year} {2019})}\BibitemShut {NoStop}%
\bibitem [{\citenamefont {Lin}\ \emph {et~al.}(2020)\citenamefont {Lin},
  \citenamefont {Calvera},\ and\ \citenamefont {Hsieh}}]{lin2020quantum}%
  \BibitemOpen
  \bibfield  {author} {\bibinfo {author} {\bibfnamefont {C.-J.}\ \bibnamefont
  {Lin}}, \bibinfo {author} {\bibfnamefont {V.}~\bibnamefont {Calvera}},\ and\
  \bibinfo {author} {\bibfnamefont {T.~H.}\ \bibnamefont {Hsieh}},\ }\bibfield
  {title} {\bibinfo {title} {Quantum many-body scar states in two-dimensional
  rydberg atom arrays},\ }\href {https://doi.org/10.1103/PhysRevB.101.220304}
  {\bibfield  {journal} {\bibinfo  {journal} {Phys. Rev. B}\ }\textbf {\bibinfo
  {volume} {101}},\ \bibinfo {pages} {220304} (\bibinfo {year}
  {2020})}\BibitemShut {NoStop}%
\bibitem [{\citenamefont {Heyl}\ \emph {et~al.}(2013)\citenamefont {Heyl},
  \citenamefont {Polkovnikov},\ and\ \citenamefont
  {Kehrein}}]{heyl2013dynamical}%
  \BibitemOpen
  \bibfield  {author} {\bibinfo {author} {\bibfnamefont {M.}~\bibnamefont
  {Heyl}}, \bibinfo {author} {\bibfnamefont {A.}~\bibnamefont {Polkovnikov}},\
  and\ \bibinfo {author} {\bibfnamefont {S.}~\bibnamefont {Kehrein}},\
  }\bibfield  {title} {\bibinfo {title} {Dynamical quantum phase transitions in
  the transverse-field ising model},\ }\href
  {https://doi.org/10.1103/PhysRevLett.110.135704} {\bibfield  {journal}
  {\bibinfo  {journal} {Phys. Rev. Lett.}\ }\textbf {\bibinfo {volume} {110}},\
  \bibinfo {pages} {135704} (\bibinfo {year} {2013})}\BibitemShut {NoStop}%
\bibitem [{\citenamefont {Cincio}\ \emph {et~al.}(2007)\citenamefont {Cincio},
  \citenamefont {Dziarmaga}, \citenamefont {Rams},\ and\ \citenamefont
  {Zurek}}]{cincio2007entropy}%
  \BibitemOpen
  \bibfield  {author} {\bibinfo {author} {\bibfnamefont {L.}~\bibnamefont
  {Cincio}}, \bibinfo {author} {\bibfnamefont {J.}~\bibnamefont {Dziarmaga}},
  \bibinfo {author} {\bibfnamefont {M.~M.}\ \bibnamefont {Rams}},\ and\
  \bibinfo {author} {\bibfnamefont {W.~H.}\ \bibnamefont {Zurek}},\ }\bibfield
  {title} {\bibinfo {title} {Entropy of entanglement and correlations induced
  by a quench: Dynamics of a quantum phase transition in the quantum ising
  model},\ }\href {https://doi.org/10.1103/PhysRevA.75.052321} {\bibfield
  {journal} {\bibinfo  {journal} {Phys. Rev. A}\ }\textbf {\bibinfo {volume}
  {75}},\ \bibinfo {pages} {052321} (\bibinfo {year} {2007})}\BibitemShut
  {NoStop}%
\bibitem [{\citenamefont {Pastori}\ \emph {et~al.}(2019)\citenamefont
  {Pastori}, \citenamefont {Heyl},\ and\ \citenamefont
  {Budich}}]{pastori2019disentangling}%
  \BibitemOpen
  \bibfield  {author} {\bibinfo {author} {\bibfnamefont {L.}~\bibnamefont
  {Pastori}}, \bibinfo {author} {\bibfnamefont {M.}~\bibnamefont {Heyl}},\ and\
  \bibinfo {author} {\bibfnamefont {J.~C.}\ \bibnamefont {Budich}},\ }\bibfield
   {title} {\bibinfo {title} {Disentangling sources of quantum entanglement in
  quench dynamics},\ }\href {https://doi.org/10.1103/PhysRevResearch.1.012007}
  {\bibfield  {journal} {\bibinfo  {journal} {Phys. Rev. Res.}\ }\textbf
  {\bibinfo {volume} {1}},\ \bibinfo {pages} {012007} (\bibinfo {year}
  {2019})}\BibitemShut {NoStop}%
\bibitem [{\citenamefont {Lee}\ \emph {et~al.}(2021)\citenamefont {Lee},
  \citenamefont {Patil}, \citenamefont {Zhang},\ and\ \citenamefont
  {Hsieh}}]{lee2021neural}%
  \BibitemOpen
  \bibfield  {author} {\bibinfo {author} {\bibfnamefont {C.~K.}\ \bibnamefont
  {Lee}}, \bibinfo {author} {\bibfnamefont {P.}~\bibnamefont {Patil}}, \bibinfo
  {author} {\bibfnamefont {S.}~\bibnamefont {Zhang}},\ and\ \bibinfo {author}
  {\bibfnamefont {C.~Y.}\ \bibnamefont {Hsieh}},\ }\bibfield  {title} {\bibinfo
  {title} {Neural-network variational quantum algorithm for simulating
  many-body dynamics},\ }\href
  {https://doi.org/10.1103/PhysRevResearch.3.023095} {\bibfield  {journal}
  {\bibinfo  {journal} {Phys. Rev. Res.}\ }\textbf {\bibinfo {volume} {3}},\
  \bibinfo {pages} {023095} (\bibinfo {year} {2021})}\BibitemShut {NoStop}%
\bibitem [{\citenamefont {Sinibaldi}\ \emph
  {et~al.}(2023{\natexlab{b}})\citenamefont {Sinibaldi}, \citenamefont
  {Giuliani}, \citenamefont {Carleo},\ and\ \citenamefont
  {Vicentini}}]{sinibaldi2023unbiasing}%
  \BibitemOpen
  \bibfield  {author} {\bibinfo {author} {\bibfnamefont {A.}~\bibnamefont
  {Sinibaldi}}, \bibinfo {author} {\bibfnamefont {C.}~\bibnamefont {Giuliani}},
  \bibinfo {author} {\bibfnamefont {G.}~\bibnamefont {Carleo}},\ and\ \bibinfo
  {author} {\bibfnamefont {F.}~\bibnamefont {Vicentini}},\ }\bibfield  {title}
  {\bibinfo {title} {Unbiasing time-dependent {V}ariational {M}onte {C}arlo by
  projected quantum evolution},\ }\href
  {https://doi.org/10.22331/q-2023-10-10-1131} {\bibfield  {journal} {\bibinfo
  {journal} {{Quantum}}\ }\textbf {\bibinfo {volume} {7}},\ \bibinfo {pages}
  {1131} (\bibinfo {year} {2023}{\natexlab{b}})}\BibitemShut {NoStop}%
\bibitem [{\citenamefont {Metropolis}\ and\ \citenamefont
  {Ulam}(1949)}]{metropolis1949}%
  \BibitemOpen
  \bibfield  {author} {\bibinfo {author} {\bibfnamefont {N.}~\bibnamefont
  {Metropolis}}\ and\ \bibinfo {author} {\bibfnamefont {S.}~\bibnamefont
  {Ulam}},\ }\bibfield  {title} {\bibinfo {title} {The monte carlo method},\
  }\href {https://doi.org/10.1080/01621459.1949.10483310} {\bibfield  {journal}
  {\bibinfo  {journal} {Journal of the American statistical association}\
  }\textbf {\bibinfo {volume} {44}},\ \bibinfo {pages} {335} (\bibinfo {year}
  {1949})}\BibitemShut {NoStop}%
\bibitem [{\citenamefont {Metropolis}\ \emph {et~al.}(1953)\citenamefont
  {Metropolis}, \citenamefont {Rosenbluth}, \citenamefont {Rosenbluth},
  \citenamefont {Teller},\ and\ \citenamefont {Teller}}]{metropolis1953}%
  \BibitemOpen
  \bibfield  {author} {\bibinfo {author} {\bibfnamefont {N.}~\bibnamefont
  {Metropolis}}, \bibinfo {author} {\bibfnamefont {A.~W.}\ \bibnamefont
  {Rosenbluth}}, \bibinfo {author} {\bibfnamefont {M.~N.}\ \bibnamefont
  {Rosenbluth}}, \bibinfo {author} {\bibfnamefont {A.~H.}\ \bibnamefont
  {Teller}},\ and\ \bibinfo {author} {\bibfnamefont {E.}~\bibnamefont
  {Teller}},\ }\bibfield  {title} {\bibinfo {title} {{Equation of State
  Calculations by Fast Computing Machines}},\ }\href
  {https://doi.org/10.1063/1.1699114} {\bibfield  {journal} {\bibinfo
  {journal} {The Journal of Chemical Physics}\ }\textbf {\bibinfo {volume}
  {21}},\ \bibinfo {pages} {1087} (\bibinfo {year} {1953})}\BibitemShut
  {NoStop}%
\bibitem [{\citenamefont {Hastings}(1970)}]{hastings1970}%
  \BibitemOpen
  \bibfield  {author} {\bibinfo {author} {\bibfnamefont {W.~K.}\ \bibnamefont
  {Hastings}},\ }\bibfield  {title} {\bibinfo {title} {{Monte Carlo sampling
  methods using Markov chains and their applications}},\ }\href
  {https://doi.org/10.1093/biomet/57.1.97} {\bibfield  {journal} {\bibinfo
  {journal} {Biometrika}\ }\textbf {\bibinfo {volume} {57}},\ \bibinfo {pages}
  {97} (\bibinfo {year} {1970})}\BibitemShut {NoStop}%
\bibitem [{Ham()}]{Hamiltonians}%
  \BibitemOpen
  \href@noop {} {}\bibinfo {howpublished} {This can be extended to
  time-dependent Hamiltonians, but it is beyond the scope of the current
  work.}\BibitemShut {Stop}%
\bibitem [{\citenamefont {Trotter}(1959)}]{trotter1959}%
  \BibitemOpen
  \bibfield  {author} {\bibinfo {author} {\bibfnamefont {H.~F.}\ \bibnamefont
  {Trotter}},\ }\bibfield  {title} {\bibinfo {title} {On the product of
  semi-groups of operators},\ }\href {http://www.jstor.org/stable/2033649}
  {\bibfield  {journal} {\bibinfo  {journal} {Proceedings of the American
  Mathematical Society}\ }\textbf {\bibinfo {volume} {10}},\ \bibinfo {pages}
  {545} (\bibinfo {year} {1959})}\BibitemShut {NoStop}%
\bibitem [{\citenamefont {Suzuki}(1976)}]{suzuki1976}%
  \BibitemOpen
  \bibfield  {author} {\bibinfo {author} {\bibfnamefont {M.}~\bibnamefont
  {Suzuki}},\ }\bibfield  {title} {\bibinfo {title} {Generalized trotter's
  formula and systematic approximants of exponential operators and inner
  derivations with applications to many-body problems},\ }\href
  {https://doi.org/10.1007/BF01609348} {\bibfield  {journal} {\bibinfo
  {journal} {Communications in Mathematical Physics}\ }\textbf {\bibinfo
  {volume} {51}},\ \bibinfo {pages} {183} (\bibinfo {year} {1976})}\BibitemShut
  {NoStop}%
\bibitem [{\citenamefont {Kingma}\ and\ \citenamefont
  {Ba}(2017)}]{kingma2014adam}%
  \BibitemOpen
  \bibfield  {author} {\bibinfo {author} {\bibfnamefont {D.~P.}\ \bibnamefont
  {Kingma}}\ and\ \bibinfo {author} {\bibfnamefont {J.}~\bibnamefont {Ba}},\
  }\href@noop {} {\bibinfo {title} {Adam: A method for stochastic
  optimization}} (\bibinfo {year} {2017}),\ \Eprint
  {https://arxiv.org/abs/1412.6980} {arXiv:1412.6980 [cs.LG]} \BibitemShut
  {NoStop}%
\bibitem [{\citenamefont {Hofmann}\ \emph {et~al.}(2022)\citenamefont
  {Hofmann}, \citenamefont {Fabiani}, \citenamefont {Mentink}, \citenamefont
  {Carleo},\ and\ \citenamefont {Sentef}}]{hofmann2022role}%
  \BibitemOpen
  \bibfield  {author} {\bibinfo {author} {\bibfnamefont {D.}~\bibnamefont
  {Hofmann}}, \bibinfo {author} {\bibfnamefont {G.}~\bibnamefont {Fabiani}},
  \bibinfo {author} {\bibfnamefont {J.~H.}\ \bibnamefont {Mentink}}, \bibinfo
  {author} {\bibfnamefont {G.}~\bibnamefont {Carleo}},\ and\ \bibinfo {author}
  {\bibfnamefont {M.~A.}\ \bibnamefont {Sentef}},\ }\bibfield  {title}
  {\bibinfo {title} {Role of stochastic noise and generalization error in the
  time propagation of neural-network quantum states},\ }\href
  {https://doi.org/10.21468/SciPostPhys.12.5.165} {\bibfield  {journal}
  {\bibinfo  {journal} {SciPost Phys.}\ }\textbf {\bibinfo {volume} {12}},\
  \bibinfo {pages} {165} (\bibinfo {year} {2022})}\BibitemShut {NoStop}%
\bibitem [{reg()}]{regularization}%
  \BibitemOpen
  \href@noop {} {}\bibinfo {howpublished} {We note however that a different
  type of regularization could be performed
  ~\cite{schmitt2020quantum}}\BibitemShut {NoStop}%
\bibitem [{nsc()}]{nscc}%
  \BibitemOpen
  \href {https://www.nscc.sg/} {}\bibinfo {howpublished}
  {\url{https://www.nscc.sg/}}\BibitemShut {NoStop}%
\end{thebibliography}%
\end{document}